\documentclass[aps,pre,onecolumn,floatfix]{revtex4}

\usepackage{bm}
\usepackage{graphicx}
\usepackage{amssymb,amsfonts,amsmath}
\usepackage{epsf}
\usepackage{hyperref}
\usepackage{subfigure}
\usepackage{epstopdf}
\usepackage{mathrsfs}
\usepackage{xcolor}
\usepackage{soul}
\usepackage{upgreek}

\usepackage{caption}

\newcommand{\be}{\begin{equation}}
\newcommand{\ee}{\end{equation}}
\newcommand{\bea}{\begin{eqnarray}}
\newcommand{\eea}{\end{eqnarray}}

\allowdisplaybreaks

\begin{document}

\title{Poles of hydrodynamic spectral functions and \\ Einstein-Helfand formulas for transport coefficients}

\author{Jo\"el Mabillard}
\email{Joel.Mabillard@ulb.be; \vfill\break ORCID: 0000-0001-6810-3709.}
\author{Pierre Gaspard}
\email{Gaspard.Pierre@ulb.be; \vfill\break ORCID: 0000-0003-3804-2110.}
\affiliation{Center for Nonlinear Phenomena and Complex Systems, Universit{\'e} Libre de Bruxelles (U.L.B.), Code Postal 231, Campus Plaine, B-1050 Brussels, Belgium}

\begin{abstract}
The local-equilibrium approach to transport processes is related to the approach based on time-dependent correlation functions and their associated spectral functions characterizing the equilibrium fluctuations of particle, momentum and other densities.  On the one hand, the transport coefficients are calculated with the Einstein-Helfand formulas derived in the local-equilibrium approach.  On the other hand, the poles of the spectral functions at complex frequencies give the damping rates of the hydrodynamic modes.  Since these rates also depend on the transport coefficients, their values can be compared to the predictions of the local-equilibrium approach.  This comparison is systematically carried out for the hard-sphere fluid by computing numerically the transport coefficients, the spectral functions, and their poles as a function of the wave number in the hydrodynamic limit.  The study shows the consistency between the two approaches for the determination of the transport properties.
\end{abstract}

\maketitle

\section{Introduction}

In the bulk phases of matter, transport properties such as viscosity and heat conduction are irreversible processes contributing to energy dissipation and entropy production \cite{GM84}.  The computation of the transport coefficients from first principles is a major challenge in statistical mechanics.  In order to predict the values of the transport coefficients, there exist several approaches based on different hypotheses.  In such a context, a general method consists in comparing the values obtained from two or more approaches in order to validate the hypotheses and the results.  Here, the local-equilibrium approach to transport theory \cite{M58,McL63,R66,P68,Z74,AP81,OL79,S14,DLW20,MG20,MG21,MG23,G22} is compared with the approach based on the correlation and spectral functions of microscopic hydrodynamics \cite{F75,BP76,BY80} to determine the transport properties in one-component fluids.

In the local-equilibrium approach, the macroscopic equations of hydrodynamics can be derived from the microscopic Hamiltonian dynamics by expressing the statistical distribution in terms of the local temperature, fluid velocity, and chemical potential, and using expansions in powers of the gradients of these macrofields.  The transport coefficients are thus given by Green-Kubo formulas or, equivalently, by Einstein-Helfand formulas \cite{G52,G54,K57,E26,H60}.  The latter generalize the Einstein formula for the diffusion coefficient of a Brownian particle \cite{E26} to all the transport coefficients by replacing the position of the random walker with the so-called Helfand moment associated to each transport property \cite{H60}.  The Einstein-Helfand formulas provide a powerful method to compute the transport coefficients, namely, the shear and bulk viscosities and the heat conductivity in the fluid phase.

Another approach is based on the time-dependent correlation functions characterizing the hydrodynamic fluctuations of particle density, momentum, and energy around equilibrium \cite{F75,BP76,BY80}.  Their temporal Fourier transforms define the associated spectral functions, which depend on the frequency $\omega$ and the wave number $q$ of the hydrodynamic modes.  The spectral function of the density fluctuations is the so-called dynamic structure factor, which determines the cross-sections for light, X-ray, and neutron scattering used to probe the properties of condensed matter since the fifties \cite{vH54}.  In these scattering processes, the interaction with a hydrodynamic mode of frequency $\omega$ and wave number $q$ generates related transfers of energy and momentum for the photon or the neutron.  Accordingly, the spectral functions present resonances corresponding to the different modes.  In fluids, there exist five hydrodynamic modes, which are associated with the five fundamental conservation laws, controlling the slowest movements in the system.  The point is that the spectral functions can be extended from real to complex frequencies in order to obtain the poles underlying the resonances.  Moreover, the complex frequencies of the poles give the dispersion relations of the hydrodynamic modes by their dependence on the wave number and the imaginary part of the complex frequencies gives the damping rate of the modes.  In the limit of small values for the wave number, these damping rates are directly related to the transport coefficients.  Since the damping rates give the widths of the resonances observed in the spectral functions, the transport coefficients can also be computed in this way by locating the poles at the complex frequencies underlying the resonances.  With this complementary approach, the values of the transport coefficients given by the Einstein-Helfand formulas can be tested numerically with molecular dynamics simulations, provided the simulated system is sufficiently large to reach the hydrodynamic regime with small enough wave numbers.

In this paper, our purpose is to use this method of comparison for the evaluation of the transport coefficients in the fluid phase of the hard-sphere system.  The dynamics of hard spheres undergoing elastic collisions can be efficiently simulated with the event-driven algorithm \cite{H97}, which is fast enough to study the hydrodynamic properties in systems with $N=500$ up to $N= 5324$ hard spheres and to validate the method of comparison.  With these numerical simulations, we compute the correlation and spectral functions for the fluctuations of particle and momentum densities.  Rational functions are fitted to the spectral functions in order to obtain numerical approximations of their poles for several small enough values of the wave number.  The dispersion relations of the hydrodynamic modes are numerically obtained in the hydrodynamic regime.  The values of their imaginary part, i.e., the widths of the resonances, can be compared with the values predicted by the transport coefficients obtained with the Einstein-Helfand formulas.

The issue is reminiscent of finding the Pollicott-Ruelle resonances \cite{P85,P86,R86a,R86b} associated with the modes of diffusion in the periodic Lorentz gas and related systems \cite{G96,GCGD01}.  In that context, the Pollicott-Ruelle resonances can be considered as generalized eigenvalues at complex frequencies for the Liouvillian dynamics of these models of deterministic diffusion.  These considerations extend from the modes of diffusion to the hydrodynamic modes, as discussed in the literature \cite{KS68,G98,G22}.  In this regard, our study also aims at relating the poles of the hydrodynamic spectral functions to the resonances of the microscopic dynamics of the fluid.

The plan of the paper is the following. In section~\ref{sec:statmech}, the statistical mechanics of fluids is summarized starting from the Hamiltonian microdynamics.  Section~\ref{sec:leq} briefly presents the local-equilibrium approach leading to the Einstein-Helfand formulas for the transport coefficients of the fluid.  The approach based on correlation and spectral functions is given in section~\ref{sec:correl+spectr-fns}, where the poles of the hydrodynamic modes are identified.  The two approaches are applied to the hard-sphere fluid in section~\ref{sec:HSF}.  The conclusion and perspectives are given in section~\ref{sec:conclusion}.

{\it Notations.} The Latin indices $a, b, c, \ldots = x, y, z$ correspond to spatial coordinates and the Greek indices $\alpha, \beta, \ldots$ label the hydrodynamic variables. Unless explicitly stated, Einstein's convention of summation over repeated indices is adopted. $\hbar$ denotes Planck's constant, $k_{\rm B}$ Boltzmann's constant, and ${\mathrm i}=\sqrt{-1}$.

\section{The statistical mechanics of fluids}
\label{sec:statmech}

\subsection{The Hamiltonian microdynamics of fluids}

At the microscopic scale, fluids are composed of atoms and/or molecules in motion according to Hamiltonian dynamics.   The microdynamics is well approximated at room temperature by the classical mechanics of the nuclei. The motion of these particles can be simulated with the technique of molecular dynamics in a finite volume $V$ with periodic boundary conditions.  If the system contains $N$ particles, the dynamics evolves in time in the $6N$-dimensional phase space of their positions ${\bf r}_i=(r_i^{x},r_i^{y},r_i^{z})$ and momenta ${\bf p}_i=(p_i^{x},p_i^{y},p_i^{z})$ with $1\leq i \leq N$.  If the spatial domain is cubic, the three components of the positions satisfy $0\leq r_i^{a} <L$, so that the volume is equal to $V=L^3$. Because of the periodic boundary conditions, this spatial domain forms a three-dimensional torus.  Every time a particle exits the cubic domain across one of its borders, the particle enters at the opposite border.  We note that the dynamics can be periodically extended into a lattice composed of images of the $N$ particles.  The minimum image convention assumes that the only image that is considered is the one in the aforementioned lattice cell \cite{H97}.

For nonrelativistic identical particles of mass $m$, the Hamiltonian function of the $N$-particle system has the following form
\be\label{eq:Hamiltonian}
H=\sum_{i=1}^N \frac{{\bf p}_i^2}{2m} + \frac{1}{2} \sum_{i\ne j} u^{(2)}(r_{ij})
\ee
in the case of binary interaction with energy potential $u^{(2)}(r)$, where $r_{ij}=\Vert{\bf r}_i-{\bf r}_j\Vert$ is the distance between the two particles $i$ and $j$.  The interaction potential is assumed to have a finite range smaller than half the size of the spatial domain, $L/2$.  Possible contributions from ternary or higher interactions may be added to this Hamiltonian function.  In the phase space of coordinates $\Gamma=\{ {\bf r}_i, {\bf p}_i\}_{i=1}^N$, the time evolution is generated by solving the Hamilton equations of motion given by
\be
\frac{{\mathrm d}{\bf r}_i}{{\mathrm d}t} = \frac{{\bf p}_i}{m}
\qquad\mbox{and}\qquad
\frac{{\mathrm d}{\bf p}_i}{{\mathrm d}t} = {\bf F}_i({\bf r}_1,{\bf r}_2,\dots,{\bf r}_N) = \sum_{j(\ne i)} {\bf F}_{ij}
\qquad\mbox{with}\qquad
{\bf F}_{ij} = - \frac{\partial u^{(2)}(r_{ij})}{\partial{\bf r}_i} \, ,
\ee
which are equivalent to Newton's equations with positional interaction forces.  The solutions of these equations form the phase-space trajectories $\Gamma(t)=\{ {\bf r}_i(t), {\bf p}_i(t)\}_{i=1}^N$ evolving in time $t$ from the initial conditions $\Gamma(0)$.  Moreover, we note that the equations of motion are symmetric under time reversal $\Theta\{ {\bf r}_i, {\bf p}_i\}=\{ {\bf r}_i, -{\bf p}_i\}$, which is a property called microreversibility.

The observable quantities are phase-space functions $A(\Gamma)$, which change in time according to $A[\Gamma(t)]$. The total mass $M=mN$, the total energy $E=H$, and the total linear momentum ${\bf P}=\sum_{i=1}^N{\bf p}_i$ are conserved by the dynamics.  The observable quantities also include the associated microscopic densities of mass, energy, and linear momentum.  The mass density can be expressed as $\hat\rho =m \hat n$ in terms of the particle density $\hat n({\bf r})=\sum_{i=1}^N \delta({\bf r}-{\bf r}_i)$.  The energy density is defined as $\hat\epsilon({\bf r}) =\sum_{i=1}^N \varepsilon_i \delta({\bf r}-{\bf r}_i)$ with $\varepsilon_i=\frac{{\bf p}_i^2}{2m} + \frac{1}{2} \sum_{j(\ne i)} u^{(2)}(r_{ij})$; and the momentum density as $\hat{g}^a({\bf r})=\sum_{i=1}^N p_i^a \delta({\bf r}-{\bf r}_i)$.  These densities $\hat c^{\alpha}=(\hat\rho,\hat\epsilon,\hat g^a)$ associated with the five fundamental conservation laws  evolve in time according to $\hat c^{\alpha}({\bf r},t) \equiv \hat c^{\alpha}[{\bf r};\Gamma(t)]$ and they obey local conservation equations of the following form,
\be
\partial_t \, \hat c^{\alpha} + \nabla^a \hat J_{c^\alpha}^a = 0 \, ,
\label{eq:local-eqs}
\ee
which are expressed in terms of corresponding current densities $\hat J_{c^\alpha}^a$.

\subsection{Statistical mechanics at equilibrium}

\subsubsection{Equilibrium probability distribution}

At thermodynamic equilibrium, matter has stationary macroscopic properties, although the particles are always in motion at the microscale because of thermal fluctuations.  For this reason, thermodynamic equilibrium is described in terms of a stationary probability distribution ${\cal P}_{\rm eq}(\Gamma)$. Depending on whether the system is isolated, in contact with a single heat reservoir, or in contact with a particle reservoir, the equilibrium probability distribution is known to be microcanonical, canonical, or grand canonical \cite{LL80a,B75}.  Here, we consider the following microcanonical probability distribution,
\be
{\cal P}_{\rm eq}(\Gamma) = \frac{1}{K(E,{\bf P})} \, \delta[E-H(\Gamma)] \; \delta\Big({\bf P}-\sum_{i=1}^{N}{\bf p}_i\Big) ,
\label{eq:microcan}
\ee
such that the mean values of the observables are given by
\be
\langle A \rangle_{\rm eq}= \frac{1}{N!} \int_{{\mathbb R}^{6N}} A(\Gamma) \, {\cal P}_{\rm eq}(\Gamma) \, {\mathrm d}\Gamma
\qquad\mbox{and}\qquad
\partial_t \langle A \rangle_{\rm eq}= 0 \, ,
\label{eq:A-equil}
\ee
where the division by $N!$ is a consequence of the indistinguishability of the particles and  $K(E,{\bf P})$ is the normalization constant such that $\frac{1}{N!} \int_{{\mathbb R}^{6N}}  {\cal P}_{\rm eq}(\Gamma)\, {\mathrm d}\Gamma =1$.

Accordingly, the $N$-particle system has fixed values for the total energy $E$ and the total linear momentum $\bf P$.
The latter is taken equal to zero ${\bf P}=0$ in the following.  Consequently, the mean momentum density is equal to zero $\langle\hat g^a\rangle_{\rm eq}=0$.  

The temperature is measured by
\be\label{eq:temperature}
\left\langle\frac{{\bf p}_i^2}{2m}\right\rangle_{\rm eq} = \frac{3}{2} \, k_{\rm B}T
\ee
for any particle $1\leq i \leq N$.  The mean particle density is given by $n=\langle\hat n\rangle_{\rm eq}=N/V$, the mean mass density by $\rho=\langle\hat\rho\rangle_{\rm eq}=mn=M/V$, and the specific volume by $v=1/\rho$.

The equilibrium properties can be computed using time averages over long enough trajectories generated by molecular dynamics at given total energy $E$ and total linear momentum ${\bf P}=0$, which should be equal to statistical averages~(\ref{eq:A-equil}) over the microcanonical probability distribution~(\ref{eq:microcan}), if the dynamics is ergodic.

\subsubsection{Equations of state for internal energy and pressure}

In the large-system limit where $N,V\to\infty$ and $n=N/V$ is kept constant, the equation of state for the specific internal energy, i.e., the internal energy per unit mass $e =\langle\hat\epsilon\rangle_{\rm eq}/\rho=E/M$ with $E=\langle H\rangle_{\rm eq}$, is  given by
\be
e(n,T) = \frac{3k_{\rm B}T}{2m} + \frac{1}{2mN} \Big\langle \sum_{i\ne j} u^{(2)}(r_{ij})\Big\rangle_{\rm eq}
\label{eq:int_energy}
\ee
and the equation of state for the hydrostatic pressure by
\be
p(n,T) = nk_{\rm B}T + \frac{1}{6V} \Big\langle \sum_{i\ne j} {\bf r}_{ij}\cdot{\bf F}_{ij}\Big\rangle_{\rm eq}
\label{eq:pressure}
\ee
with ${\bf r}_{ij}={\bf r}_i-{\bf r}_j$.

All the equilibrium properties can be deduced from these two equations of state, including the specific entropy $s$ such that ${\mathrm d}e=T{\mathrm d}s-p{\mathrm d}v$, the specific enthalpy $h=e+pv$, the specific heat capacities at constant specific volume and at constant pressure
\be
c_v = \left(\frac{\partial e}{\partial T}\right)_v = T  \left(\frac{\partial s}{\partial T}\right)_v
\qquad\mbox{and}\qquad
c_p = \left(\frac{\partial h}{\partial T}\right)_p = T  \left(\frac{\partial s}{\partial T}\right)_p \, ,
\ee
the adiabatic and isothermal compressibilities
\be
\chi_s = \frac{1}{\rho}  \left(\frac{\partial \rho}{\partial p}\right)_s 
\qquad\mbox{and}\qquad
\chi_T = \frac{1}{\rho}  \left(\frac{\partial \rho}{\partial p}\right)_T \, ,
\ee
and the related adiabatic and isothermal speeds of sound
\be
\label{eq:sound_speeds}
c_s = \sqrt{\left(\frac{\partial p}{\partial \rho}\right)_s}
\qquad\mbox{and}\qquad
c_T = \sqrt{\left(\frac{\partial p}{\partial \rho}\right)_T} \, .
\ee
These pairs of derived quantities have the common ratio:
\be
\label{eq:ratio}
\gamma \equiv \frac{c_p}{c_v} = \frac{\chi_T}{\chi_s} = \frac{c_s^2}{c_T^2} \, .
\ee

\subsubsection{Static structure factor}

Fluids have isotropic and uniform equilibrium properties on large scales.  Nevertheless, on small scales, there may exist nontrivial statistical correlations between the positions of the particles at a given time because of their mutual interaction.  These correlations manifest themselves in the fluctuations of the microscopic particle density $\hat n({\bf r},t)$ and they can be characterized by the {\it static structure factor} defined as~\cite{BP76, BY80}
\begin{align}
\label{eq:ssf}
S(q)\equiv \frac{1}{N}\left\langle \delta \hat n({\bf q},0)\, \delta \hat n^*({\bf q},0) \right\rangle_{\rm eq}=\frac{1}{N}\left\langle \sum_{i,j=1}^N {\rm e}^{{\mathrm i}{\bf q}\cdot\left[{\bf r}_i(0)-{\bf r}_j(0)\right]} \right\rangle_{\rm eq} - n \, (2\pi)^3 \, \delta({\bf q})\, ,
\end{align}
where $\delta\hat n({\bf q},t)=\int_V \delta\hat n({\bf r},t){\rm e}^{{\mathrm i}{\bf q}\cdot{\bf r}}{\mathrm d}{\bf r}$ is the spatial Fourier transform of the density fluctuations with respect to the mean density: $\delta\hat n({\bf r},t)\equiv \hat n({\bf r},t)-\langle\hat n({\bf r})\rangle_{\rm eq}$.  The correlation function~(\ref{eq:ssf}) is an equilibrium property because it is defined at equal time $t=0$ for the two observables.  Since fluids are isotropic, the static structure factor only depends on the magnitude $q=\Vert{\bf q}\Vert$ of the wave vector ${\bf q}=(q^x,q^y,q^z)$.  We note that the static structure factor is related to the spatial Fourier transform of the pair distribution function $g(r)$ between two particles separated by the distance~$r$ at a given time \cite{BP76,BY80,B75,RD77,R98}.

In the limit of small $q$, the static structure factor gives the isothermal compressibility according to
\begin{align}
\chi_T = \lim_{q\rightarrow 0} \frac{S(q)}{n k_{\rm B} T}\, . \label{eq:isocompSq}
\end{align}

\subsection{Statistical mechanics out of equilibrium}

\subsubsection{Nonequilibrium probability distribution}

In general, the phase-space probability distribution is not stationary and its time evolution is ruled by Liouville's equation
\begin{align}
\partial_t \, {\cal P}(\Gamma,t) = \{ H(\Gamma), {\cal P}(\Gamma,t)\} \, ,
\label{eq:Liouville}
\end{align}
where $\{A,B\}$ denotes the Poisson bracket between the phase-space functions $A(\Gamma)$ and $B(\Gamma)$ \cite{B75}.  Time-dependent probability distributions such as ${\cal P}(\Gamma,t)$ provide the statistical description of nonequilibrium systems evolving in time from arbitrary initial conditions given by ${\cal P}(\Gamma,0)$.

\subsubsection{Towards the macroscopic hydrodynamic equations}

The nonequilibrium mean values are defined as
\begin{align}
\langle\hat c^{\alpha}({\bf r})\rangle_t \equiv \frac{1}{N!} \int_{{\mathbb R}^{6N}} \hat c^{\alpha}({\bf r};\Gamma) \, {\cal P}(\Gamma,t) \, {\mathrm d}\Gamma
\label{eq:neq-mean-value}
\end{align}
and they obey the mean local conservation equations
\be
\partial_t \, \langle\hat c^{\alpha}\rangle_t + \nabla^a \langle\hat J_{c^\alpha}^a\rangle_t = 0 \, ,
\label{eq:av-local-eqs}
\ee
as a consequence of the microscopic equations (\ref{eq:local-eqs}).  In the long-time limit, the nonequilibrium mean values~(\ref{eq:neq-mean-value}) are expected to converge towards their equilibrium mean values if the dynamics is mixing in phase space: $\lim_{t\to\infty} \langle\hat c^{\alpha}\rangle_t = \langle\hat c^{\alpha}\rangle_{\rm eq}$.  The relaxation to equilibrium can thus be understood on this ground.

On macroscopic scales, the mean local conservation equations~(\ref{eq:av-local-eqs}) should lead to the hydrodynamic equations including the continuity equation, the Navier-Stokes equations, and the heat equation.  This relationship is established by introducing the velocity field as the velocity of the center of mass for every fluid element as
\begin{align}
v^a({\bf r},t) \equiv \frac{\langle \hat g^a({\bf r})\rangle_t}{\langle \hat \rho({\bf r})\rangle_t} \, .
\label{eq:velocity}
\end{align}
Therefore, taking the mean value of the microscopic local conservation equation for mass $\partial_t\hat\rho+\nabla^a \hat g^a=0$ gives the continuity equation $\partial_t\rho+\nabla^a(\rho v^a)=0$ with $\rho({\bf r},t)=\langle\hat\rho({\bf r})\rangle_t$.
In this way, the nonequilibrium mean values~(\ref{eq:neq-mean-value}) can be considered as the corresponding macrofields $c^{\alpha}({\bf r},t)$ and the five macroscopic equations of hydrodynamics can be deduced within the framework of the local-equilibrium approach, as explained in the following section~\ref{sec:leq}.  Furthermore, Liouville's equation~(\ref{eq:Liouville}) also rules the time evolution of the time-dependent correlation functions and the corresponding spectral functions, which are introduced in section~\ref{sec:correl+spectr-fns}.

\section{The approach based on local equilibrium}
\label{sec:leq}

Starting from arbitrary nonequilibrium initial conditions, fluids undergo a rapid thermalization on the intercollisional time scale, during which the momentum distribution converges towards a local Maxwellian equilibrium distribution at the local temperature $T({\bf r},t)$ in the frame moving with the fluid elements at the velocity~(\ref{eq:velocity}).   Such a nonequilibrium probability distribution has the following local-equilibrium form:
\begin{align}\label{eq:leq-prob}
{\cal P}_{\rm leq}(\Gamma;\boldsymbol{\lambda}) = \exp\left[ -\Omega(\boldsymbol{\lambda})-\int_V \lambda^\alpha({\bf r}) \, \hat c^\alpha({\bf r};\Gamma) \, {\mathrm d}{\bf r} \right] ,
\end{align}
where $\Omega(\boldsymbol{\lambda})\equiv \ln\left\{\frac{1}{N!} \int_{{\mathbb R}^{6N}}  \exp[-\int_V \lambda^\alpha({\bf r}) \, \hat c^\alpha({\bf r};\Gamma) \, {\mathrm d}{\bf r} ]\, {\mathrm d}\Gamma\right\}$. Therein, the conjugate fields $\lambda^\alpha({\bf r})$ associated with the different densities can be considered as $\lambda_\rho=-\beta\mu$, $\lambda_\epsilon=\beta$, and $\lambda_{g^a}=-\beta v^a$ up to terms of second order in the gradients, $\beta=(k_{\rm B}T)^{-1}$ being the inverse temperature, $\mu$ the chemical potential, and $v^a$ the velocity field \cite{M58,McL63,R66,P68,Z74,AP81,OL79,S14,DLW20,MG20,MG21,MG23,G22}.

At any time $t$, the conjugate fields $\lambda^\alpha_t({\bf r})=\lambda^\alpha({\bf r},t)$ are defined in such a way that the mean values of the microscopic densities should always be equal to their mean values with respect to the local-equilibrium probability distribution~(\ref{eq:leq-prob}) also at time $t$: $\langle\hat c^{\alpha}\rangle_t=\langle\hat c^{\alpha}\rangle_{{\rm leq},\boldsymbol{\lambda}_t}$.  Even if the initial probability distribution is taken as a local-equilibrium one, the local-equilibrium probability distribution ${\cal P}_{\rm leq}(\Gamma;\boldsymbol{\lambda}_t)$ at time $t\ne 0$ differs in general from the exact one ${\cal P}(\Gamma,t)$, which is the solution of Liouville's equation~(\ref{eq:Liouville}), by a factor depending on the phase-space variables \cite{McL63,S14}:
\begin{align}
{\cal P}(\Gamma,t) = {\cal P}_{\rm leq}(\Gamma;\boldsymbol{\lambda}_t) \, {\rm e}^{\Sigma(\Gamma,t)} \, .
\label{eq:leq-prob-t}
\end{align}
Using an expansion of $\Sigma(\Gamma,t)$ in powers of the gradients of the macrofields, the macroscopic equations of hydrodynamics can be derived.  In this approach, the transport coefficients are computed from the microdynamics using Green-Kubo formulas \cite{G52,G54,K57}.  Accordingly, any transport coefficient is given in the large-system limit by
 \begin{align}
{\cal L}^{a}_{\alpha} = \frac{1}{V} \int_0^{\infty} \left\langle  \delta{\mathbb J}_{c^\alpha}^{a}(t) \, \delta{{\mathbb J}}^{a}_{c^\alpha}(0)\right\rangle_{\rm eq} \, {\mathrm d}t \, ,
\label{eq:GK-formula}
\end{align}
as the time integral of the autocorrelation function of the microscopic global current $\delta{\mathbb J}_{c^\alpha}^{a}(t)={\mathbb J}_{c^\alpha}^{a}(t)-\langle{\mathbb J}_{c^\alpha}^{a}\rangle_{\rm eq}$, which is associated with the transported quantity (and there  here is no Einstein's summation over the indices $a$ and $\alpha$).  The autocorrelation function is here calculated with respect to the microcanonical equilibrium distribution~(\ref{eq:microcan}), where there is no fluctuation of total linear momentum, energy, and particle number.  Accordingly, these global currents for the transports of linear momentum and energy are respectively given by~\cite{Z65}
\begin{align}
{\mathbb J}^{ab} &=\int_V \hat{J}^a_{g^b}({\bf r}) \, {\mathrm d}{\bf r}=\sum_{i} \frac{ p_i^{a}}{m} \, p_i^{b} + \frac{1}{2} \sum_{i\neq j} r^a_{ij} \, F_{ij}^{b} \, ,\label{eq:Jab}\\
{\mathbb J}_{\epsilon}^{a} &=\int_V \hat{J}^a_{\epsilon}({\bf r}) \, {\mathrm d}{\bf r}=  \sum_{i} \frac{{p}^a_i}{m} \, \varepsilon_{i} + \frac{1}{2} \sum_{i\neq j}r^a_{ij} \, {F}^{b}_{ij} \, \frac{{p}^b_{i}+{p}^b_{j}}{2m} \, .\label{eq:Jaq}
\end{align}
We note that the hydrostatic pressure~(\ref{eq:pressure}) can be obtained from the equilibrium mean values of the global momentum current according to $\langle{\mathbb J}^{ab}\rangle_{\rm eq} = p V \delta^{ab}$, while $\langle{\mathbb J}_{\epsilon}^{a}\rangle_{\rm eq} =0$.

Equivalently, the transport coefficients can be obtained in the combined large-system and long-time limits using the Einstein-Helfand formulas \cite{E26,H60}
 \begin{align}
{\cal L}^{a}_{\alpha} = \lim_{t\to\infty}\frac{1}{2tV} \, \left\langle\left[\Delta{\mathbb G}_{c^\alpha}^{a}(t) - \langle\Delta{\mathbb G}_{c^\alpha}^{a}(t)\rangle_{\rm eq}\right]^2\right\rangle_{\rm eq} \, ,
\label{eq:EH-formula}
\end{align}
expressed in terms of the associated Helfand moment \cite{H60}:
\begin{align}
\label{eq:Helfand}
\Delta{\mathbb G}_{c^\alpha}^{a}(t) \equiv \int_0^t {\mathbb J}_{c^\alpha}^{a}(t') \, {\mathrm d}t' \, .
\end{align}
The key point is that this Helfand moment performs a random walk and the corresponding transport coefficient is given by the diffusivity of this random walk.  This method is very powerful to compute the transport coefficients using molecular dynamics  simulations.  For this purpose, the minimum image convention should be taken into account to satisfy the periodic boundary conditions of the microdynamics \cite{BS96,VG03}.  Accordingly, the random walk of the Helfand moments can be rigorously implemented in the simulation, leading to accurate evaluations of the transport coefficients \cite{VSG07a,VSG07b}.

The shear viscosity, the bulk viscosity, and the heat conductivity are thus respectively given by
\begin{align}
\eta&=\lim_{t\rightarrow\infty}\frac{1}{2t k_{\rm B}TV}\left\langle \left[\Delta{\mathbb G}^{xy}(t)\right]^2\right\rangle_{\rm eq}\, ,\label{eq:etaH}\\
\zeta &=\lim_{t\rightarrow\infty}\frac{1}{2t k_{\rm B}TV}\left\langle \left[\Delta{\mathbb G}_{\zeta}(t)-pVt\right]^2\right\rangle_{\rm eq} \qquad\mbox{with}\qquad
\Delta{\mathbb G}_{\zeta} \equiv \frac{1}{3} (\Delta{\mathbb G}^{xx}+\Delta{\mathbb G}^{yy}+\Delta{\mathbb G}^{zz}) \, ,\label{eq:zetaH}\\
\kappa &=\lim_{t\rightarrow\infty}\frac{1}{2t k_{\rm B}T^2V} \left\langle \left[\Delta{\mathbb G}_{\epsilon}^x(t)\right]^2\right\rangle_{\rm eq}\, ,\label{eq:kappaH}
\end{align}
since $\left\langle\Delta{\mathbb G}^{xy}(t)\right\rangle_{\rm eq}=0$, $\left\langle\Delta{\mathbb G}_{\zeta}(t)\right\rangle_{\rm eq} =pVt$, and $\left\langle\Delta{\mathbb G}_{\epsilon}^x(t)\right\rangle_{\rm eq}=0$.  In this regard, the hydrostatic pressure~(\ref{eq:pressure}) can  be computed according to $p=\lim_{t\to\infty} \left\langle\Delta{\mathbb G}_{\zeta}(t)\right\rangle_{\rm eq} /(tV)$ in terms of the mean drift velocity of the Helfand moment defined in equation~(\ref{eq:zetaH}).  
We also note that the longitudinal viscosity can be directly obtained from
\begin{align}
\zeta+\frac{4}{3}\, \eta &=\lim_{t\rightarrow\infty}\frac{1}{2t k_{\rm B}TV}\left\langle \left[\Delta{\mathbb G}^{xx}(t)-pVt\right]^2\right\rangle_{\rm eq}\, .\label{eq:longviscH}
\end{align}
In equations~(\ref{eq:etaH})-(\ref{eq:longviscH}), the factors $(k_{\rm B}T)^{-1}$ and $(k_{\rm B}T^2)^{-1}$ arise from the macroscopic definitions of these transport coefficients, as compared to their generic form~(\ref{eq:EH-formula}).

Now, the issue is to test numerically these predictions by comparing them with the molecular dynamics simulation of the hydrodynamic behavior.  For this purpose, we consider the approach based on the time-dependent correlation functions of the hydrodynamic fluctuations at wave vector $\bf q$ and their associated spectral functions, as carried out in the following section~\ref{sec:correl+spectr-fns}.

\section{The approach based on correlation and spectral functions}
\label{sec:correl+spectr-fns}

\subsection{Generalities}

Nonequilibrium properties like the transport coefficients are related to time-dependent correlation functions defined with respect to the equilibrium probability distribution as
\begin{align}
C_{AB}(t) \equiv \langle \delta A(0) \, \delta B(t)\rangle_{\rm eq}
\qquad\mbox{with}\qquad
\delta A \equiv A - \langle A\rangle_{\rm eq}
\qquad\mbox{and}\qquad
\delta B \equiv B - \langle B\rangle_{\rm eq}
\end{align}
for some observable quantities $A(\Gamma)$ and $B(\Gamma)$ \cite{F75}.  If the dynamics is mixing, the memory of the initial fluctuations is lost as the time interval $t$ increases, so that the correlation functions converge to zero in the long-time limit: $\lim_{t\to\infty} C_{AB}(t)=0$.  The temporal Fourier transform of the correlation function defines the associated spectral function
\begin{align}
S_{AB}(\omega) \equiv \int_{-\infty}^{+\infty} C_{AB}(t) \, {\rm e}^{-{\mathrm i}\omega t} \, {\mathrm d}t \, ,
\end{align}
giving the frequency content of the fluctuations.

These spectral functions can be extended towards complex frequencies, where they may have poles or other possible singularities such as branch cuts.  The poles may be located at the complex frequencies $\omega_r={\rm Re}\, \omega_r + {\mathrm i}\,{\rm Im}\, \omega_r$, where the real part ${\rm Re}\, \omega_r$ gives the characteristic frequency of the corresponding mode and the imaginary part ${\rm Im}\, \omega_r$ the damping rate of the mode and, thus, its relaxation time $\tau_r=1/\vert{\rm Im}\, \omega_r\vert$.  In this way, the characteristic time scales of the fluid can be obtained as intrinsic properties of the underlying microscopic dynamics, if the locations of the poles converge in the large-system limit where $N,V\to\infty$ with a constant density $n=N/V$.

Similarly, we may consider the Laplace transform of the correlation function:
\begin{align}
\tilde C_{AB}(z) \equiv \int_{0}^{\infty} C_{AB}(t) \, {\rm e}^{-z t} \, {\mathrm d}t 
\end{align}
with $z={\mathrm i}\omega$.  The poles of the Laplace transform and the spectral function are thus related by $z_r={\mathrm i}\omega_r$.  For the dynamical system of $N$ particles moving on the torus of volume $V=L^3$, these poles may be considered as the Pollicott-Ruelle resonances of the dynamics, if they are well defined \cite{P85,P86,R86a,R86b}.

If the observables $A$ and $B$ are the Fourier modes of wave vector $\bf q$ in the fluid, the poles of the spectral functions should provide the dispersion relations $\omega_r(q)$ of the hydrodynamic modes including their relaxation rate ${\rm Im}\, \omega_r(q)$.  These dispersion relations should allow us to determine if a mode is diffusive or propagative depending on whether  the real part ${\rm Re}\, \omega_r(q)$ is equal to zero or not.  In the latter case, the propagation speed is given by $c_s=\lim_{q\to 0} \vert {\rm Re}\, \omega_r(q)\vert/q$.

\subsection{Dynamics of density fluctuations}

The time evolution of density fluctuations $\delta\hat n({\bf q},t)$ around equilibrium can be characterized by the so-called {\it intermediate scattering function} \cite{vH54}
\begin{align}
\label{eq:isf}
F(q,t) \equiv\frac{1}{N} \left\langle \delta \hat n({\bf q},t)\, \delta \hat n^*({\bf q},0) \right\rangle_{\rm eq}=\frac{1}{N}\left\langle \sum_{i,j=1}^N{\rm e}^{{\mathrm i}{\bf q}\cdot\left[{\bf r}_i(t)-{\bf r}_j(0)\right]} \right\rangle_{\rm eq} - n \, (2\pi)^3 \, \delta({\bf q}) \, ,
\end{align}
such that the static structure factor~(\ref{eq:ssf}) is recovered at time $t=0$: $S(q)=F(q,0)$.
The temporal Fourier transform of the function~(\ref{eq:isf}) defines the {\it dynamic structure factor} \cite{BY80}
\begin{align}
\label{eq:dsf}
S(q,\omega) \equiv \int_{-\infty}^{+\infty} F(q,t)\, {\rm e}^{-{\mathrm i}\omega t}\, {\mathrm d}t \, .
\end{align}
This spectral function is related to the cross-section for light, X-ray, or neutron scattering with the transfers of momentum $\hbar{\bf q}$ and energy $\hbar\omega$ \cite{vH54,F75,BP76,BY80}.  The dynamic and static structure factors are related to each other by the following sum rule,
\begin{align}
\label{eq:sr_dsf}
\frac{1}{2\pi} \int_{-\infty}^{+\infty} S(q,\omega) \, {\mathrm d}\omega = S(q)\, .
\end{align}

\subsection{Dynamics of momentum fluctuations}

Around equilibrium, the momentum fluctuations $\hat g^a({\bf r},t)$ are proportional to those of the velocity field, if equation~(\ref{eq:velocity}) is considered at the mesoscopic scale: $\hat v^a({\bf r},t)\simeq \hat g^a({\bf r},t)/\rho$.  These fluctuations can be characterized by the following time-dependent correlation functions,
\begin{align}
\label{eq:vcf}
C^{ab}(q,t)\equiv\frac{1}{N m^2}\langle\hat g^a({\bf q},t)\, \hat g^{b*}({\bf q},0)\rangle_{\rm eq} =\frac{1}{Nm^2} \left\langle \sum_{i,j=1}^N p_{i}^a(t)\, p_{j}^b(0) \, {\rm e}^{{\mathrm i}{\bf q}\cdot\left[{\bf r}_i(t)-{\bf r}_j(0)\right]} \right\rangle_{\rm eq} \, , 
\end{align}
where
\begin{align}
\label{eq:g(q)}
\hat g^a({\bf q},t) \equiv \int_V \hat g^a({\bf r},t) \, {\rm e}^{{\mathrm i}{\bf q}\cdot{\bf r}} \, {\mathrm d}{\bf r} = \sum_{i=1}^N p_i^a(t) \, {\rm e}^{{\mathrm i}{\bf q}\cdot{\bf r}_i(t)}
\end{align}
and $\delta \hat{g}^a=\hat{g}^a$ since $\langle\hat g^a\rangle_{\rm eq}=0$.
Again, these correlation functions only depend on $q=\Vert{\bf q}\Vert$, because the fluid is isotropic.

The associated spectral function is defined as
\begin{align}
\label{eq:vsf}
J^{ab}(q,\omega) \equiv \int_{-\infty}^{+\infty} C^{ab}(q,t)\, {\rm e}^{-{\mathrm i}\omega t}\, {\mathrm d}t \, ,
\end{align}
which obeys the sum rule \cite{BY80}
\begin{align}
\label{eq:sr_vsf}
\frac{1}{2\pi} \int_{-\infty}^{+\infty} J^{ab}(q,\omega) \, {\mathrm d}\omega = C^{ab}(q,0) = \frac{k_{\rm B}T}{m} \, \delta^{ab} \, .
\end{align}

A further consequence of isotropy in fluids is that the tensorial correlation and spectral functions (\ref{eq:vcf}) and (\ref{eq:vsf}) can be decomposed into their longitudinal and transverse components as
\begin{align}
\label{eq:vcf+vsf-l+t}
C^{ab}(q,t) & = \frac{q^a q^b}{q^2} \, C_{\rm l}(q,t) + \left( \delta^{ab} - \frac{q^a q^b}{q^2}\right) \, C_{\rm t}(q,t) \, , \\
J^{ab}(q,\omega) & = \frac{q^a q^b}{q^2} \, J_{\rm l}(q,\omega) + \left( \delta^{ab} - \frac{q^a q^b}{q^2}\right) \, J_{\rm t}(q,\omega) \, .
\end{align}
Equation~(\ref{eq:sr_vsf}) implies that $C_{\rm l}(q,0)=C_{\rm t}(q,0) = k_{\rm B}T/m$ and the other sum rules,
\begin{align}
\label{eq:sr_vsf-l+t}
\frac{1}{2\pi} \int_{-\infty}^{+\infty} J_{\rm l}(q,\omega) \, {\mathrm d}\omega = \frac{1}{2\pi} \int_{-\infty}^{+\infty} J_{\rm t}(q,\omega) \, {\mathrm d}\omega = \frac{k_{\rm B}T}{m} \, .
\end{align}

Remarkably, the longitudinal momentum correlation function is related to the intermediate scattering function~(\ref{eq:isf}) and the associated longitudinal spectral function to the dynamic structure factor according to
\begin{align}
\label{eq:lca_hydro}
C_{\rm l}(q,t) = -\frac{1}{q^2}\frac{{\mathrm d}^2}{{\mathrm d}t^2}F(q,t) 
\qquad\mbox{and}\qquad
J_{\rm l}(q,\omega)=\frac{\omega^2}{q^2}S(q,\omega)\, .
\end{align}
This result is the consequence of the property that
\begin{align}
\frac{{\mathrm d}}{{\mathrm d}t} \langle \delta A(0) \, \delta B(t)\rangle_{\rm eq} = \langle \delta A(0) \, \frac{{\mathrm d}}{{\mathrm d}t} \delta B(t)\rangle_{\rm eq} = - \langle \left[\frac{{\mathrm d}}{{\mathrm d}t}\delta A(t)\right]_{t=0} \, \delta B(t)\rangle_{\rm eq} \, ,
\end{align}
resulting from the stationarity of the equilibrium probability distribution.  Accordingly, the longitudinal component of the momentum correlation and spectral functions can be directly obtained from the previously defined correlation and spectral functions characterizing the density fluctuations.

\subsection{Hydrodynamic behavior of the correlation and spectral functions}

\begin{figure}[h!]\centering
{\includegraphics[width=0.8\textwidth]{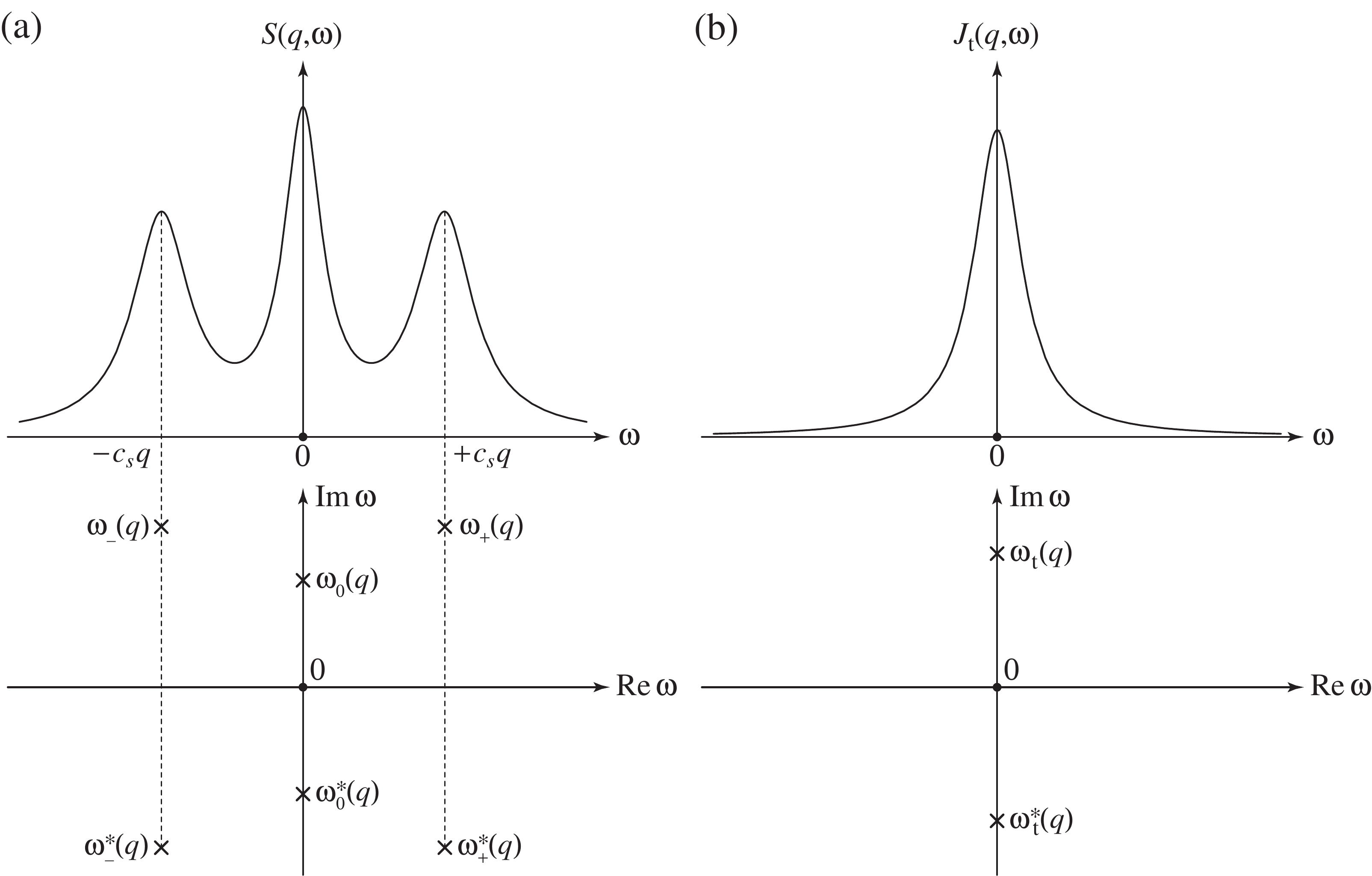}}
\caption[] {Schematic representation of the dynamic structure factor $S(q,\omega)$ and the transverse component of the spectral function of momentum fluctuations versus frequency $\omega$ and their underlying poles depicted in the plane of complex frequencies.}\label{Fig:Poles}
\end{figure}

On large spatiotemporal scales, the system is expected to relax towards thermodynamic equilibrium and the time-dependent correlation functions to decay in time according to macroscopic hydrodynamics linearized around equilibrium, as supposed by Onsager's hypothesis of regression of fluctuations \cite{O31b}.  This hypothesis is at the basis of fluctuating hydrodynamics \cite{LL57,LL80b,OS06}.  Moreover, it can be justified in the framework of the local-equilibrium approach as shown in appendix~\ref{app:rh-leq}.  Accordingly, the correlation and spectral functions can be calculated in the hydrodynamic regime by solving the linearized hydrodynamic equations including the contributions from the viscosities and heat conduction, as summarized in appendix~\ref{app:hydro-cf+sf}.  Thus, the hydrodynamic approximations of the spectral functions are given by \cite{BP76,BY80}
\begin{align}
\label{eq:dsf_hydro}
\frac{S(q,\omega)}{S(q)} & = \left(1-\frac{1}{\gamma}\right)\frac{2D_Tq^2}{\omega^2+(D_Tq^2)^2}+\frac{1}{\gamma}\left[\frac{\Gamma q^2}{(\omega+c_sq)^2+(\Gamma q^2)^2}+\frac{\Gamma q^2}{(\omega-c_sq)^2+(\Gamma q^2)^2}\right]\notag\\
&\qquad +\frac{3\Gamma-D_v}{\gamma \, c_s} q \left[\frac{\omega+c_sq}{(\omega+c_sq)^2+(\Gamma q^2)^2}-\frac{\omega-c_sq}{(\omega-c_sq)^2+(\Gamma q^2)^2}\right] ,
\end{align}
\begin{align}
\label{eq:tvsf_hydro}
\frac{J_{\rm t}(q,\omega)}{C_{\rm t}(q,0)} = \frac{2\nu q^2}{\omega^2+\left(\nu q^2\right)^2} \, ,
\end{align}
and $J_{\rm l}(q,\omega)$ can be deduced from equation~(\ref{eq:dsf_hydro}) with equation~(\ref{eq:lca_hydro}), where $c_s$ is the adiabatic (i.e., isoentropic) speed of sound introduced in equation~(\ref{eq:sound_speeds}),  $\gamma$ the ratio~(\ref{eq:ratio}), $D_T=\kappa/(\rho c_p)$ the thermal diffusivity, $D_v=(\zeta+\frac{4}{3}\eta)/\rho$ the longitudinal kinematic viscosity, $\Gamma\equiv\frac{1}{2}\left[D_v+D_T(\gamma-1)\right]$ the acoustic attenuation coefficient, and $\nu\equiv \eta/\rho$ the transverse kinematic viscosity.  The spectral functions are schematically represented in figure~\ref{Fig:Poles}.

In the hydrodynamic regime, the spectral functions present poles located at the following complex frequencies:
\begin{align}
\omega_0(q) &= {\mathrm i} \, D_T q^2 + \cdots\, , \label{eq:heat_pole} \\
\omega_{\pm}(q) &= \pm c_s q + {\mathrm i} \, \Gamma q^2 + \cdots \, , \label{eq:sound_poles} \\
\omega_{\rm t}(q) &= {\mathrm i} \, \nu q^2 + \cdots \, , \label{eq:shear_pole} 
\end{align}
and their complex conjugates $\omega_0^*(q)$, $\omega_{\pm}^*(q)$, and $\omega_{\rm t}^*(q)$.
These complex frequencies represent the dispersion relations of the five hydrodynamic modes of the fluid, which are the diffusive heat modes, the propagative sound modes, and the diffusive shear modes, respectively. The shear modes have a multiplicity equal to two because there are two directions transverse to the wave vector $\bf q$.    In equations~(\ref{eq:heat_pole})-(\ref{eq:shear_pole}), the dots denote possible corrections vanishing faster than $q^2$ in the limit $q\to 0$.   In this regard, we note that the time-dependent correlation functions in the Green-Kubo formulas~(\ref{eq:GK-formula}) are known to manifest algebraic decays as $1/t^{3/2}$ in arbitrarily large systems and, as a consequence of these so-called long time tails, the dispersion relations of the sound modes have been shown to include a correction going as $q^{5/2}$ \cite{ED72,ED75,DvBK21}.


The location of all these poles determines the shape of the spectral functions, as shown in figure~\ref{Fig:Poles}.  In particular, the poles at the complex frequencies~(\ref{eq:heat_pole}) and~(\ref{eq:sound_poles}) and their complex conjugates  explain that the plot of the dynamic structure factor $S(q,\omega)$ as a function of the frequency $\omega$ has three resonances corresponding to the three resonances observed in figure~\ref{Fig:Poles}(a): the Rayleigh resonance around $\omega=0$ and the two Brillouin resonances around $\omega=\pm c_s q$. Similarly, the transverse component of the spectral function of momentum fluctuations has a single resonance in relation to the poles at the complex frequencies $\omega_{\rm t}(q)$ and $\omega_{\rm t}^*(q)$, as seen in figure~\ref{Fig:Poles}(b).  The widths of these resonances are given by the imaginary part of the corresponding complex frequencies.  

The important result for our purposes is that the poles of the spectral functions should allow us to obtain a further evaluation of the transport coefficients beyond their values given by the Einstein-Helfand formulas (\ref{eq:etaH})-(\ref{eq:longviscH}) and to test numerically the consistency of the predictions.  With this aim, rational functions of the form
\begin{align}\label{eq:rational_fn}
R(\omega) = \frac{\sum_{k=0}^N b_k\, \omega^k}{1+\sum_{k=1}^M a_k\, \omega^k}
\end{align}
are fitted to the spectral functions for different values of the wave number $q$.  Approximate locations for the poles can thus be determined from the zeros of the denominator.  Three pairs of poles are obtained for the dynamic structure factor and one pair for the transverse component of the spectral function of the momentum fluctuations, giving $q$-dependent values for the speed of sound and the diffusivities as
\begin{align}\label{eq:q-coefficients}
c_s(q) \equiv {\rm Re}[\omega_+(q)]/q \, , \qquad
\Gamma(q) \equiv {\rm Im}[\omega_{\pm}(q)]/q^2 \, , \qquad
D_T(q) \equiv {\rm Im}[\omega_0(q)]/q^2 \, , \qquad\mbox{and} \qquad
\nu(q) \equiv {\rm Im}[\omega_{\rm t}(q)]/q^2 \, .
\end{align}
The extrapolation of these values to $q=0$ can be compared to the values obtained using the Einstein-Helfand formulas (\ref{eq:etaH})-(\ref{eq:longviscH}).

The results of this comparison between the poles of the spectral functions and the values obtained with the Einstein-Helfand formulas are presented in the following section~\ref{sec:HSF} for the hard-sphere fluid.

\section{Results for the hard-sphere fluid}
\label{sec:HSF}

\subsection{The dynamics of the hard-sphere system}

In order to test numerically the approaches based on local equilibrium and on hydrodynamic spectral functions for computing the transport properties, we consider a dynamical system of $N$ hard spheres of diameter $d$ and mass $m$ moving in a cubic domain of volume $V=L^3$ with periodic boundary conditions.  For this system, the binary energy potential is given by
\begin{align}\label{eq:u(2)-HS}
u^{(2)}({r}_{ij})&=
\begin{cases} 
0 \qquad \text{for} \quad {r}_{ij}>d \, ,\\ 
\infty \quad\ \, \text{for} \quad {r}_{ij}\leq d \, , \\
\end{cases}
\end{align}
so that the Hamiltonian function~(\ref{eq:Hamiltonian}) reduces to the total kinetic energy if the hard spheres do not overlap, i.e., if their positions satisfy the conditions $r_{ij} >d$ for all $1\leq i \neq j\leq N$.  Therefore, the total energy and the total linear momentum that are conserved by the hard-sphere dynamics read
\begin{align}\label{eq:HS-csts-motion}
H=\sum_{i=1}^N \frac{{\bf p}_i^2}{2m}
\qquad\mbox{and}\qquad
{\bf P} = \sum_{i=1}^N {\bf p}_i \, .
\end{align}

As a consequence, the particles move in free flights interrupted by instantaneous elastic collisions occurring at successive times $\{t_c\}$ between pairs of hard spheres. During the free flights, the positions evolve in time according to 
\begin{align}\label{eq:free_flight}
{\bf r}_i(t) = {\bf r}_i(t_c) + \frac{{\bf p}_i(t_c+0)}{m} \, (t-t_c) 
\qquad\mbox{for}\qquad
t_c < t< t_{c+1} \, .
\end{align} Upon an elastic collision between the hard spheres $i$ and $j$, the positions of the two particles remain continuous, so that ${\bf r}_{i}(t_c+0)={\bf r}_{i}(t_c-0)={\bf r}_{i}(t_c)\equiv {\bf r}_{i}^{(c)}$, but their momenta are subject to the following discontinuous changes:
\begin{align}\label{eq:elastic_coll}
\left\{
\begin{array}{l}
{\bf p}_i(t_c+0) = {\bf p}_i(t_c-0) - \frac{2}{d^2} \big[ {\bf r}_{ij}^{(c)}\cdot {\bf p}_{ij}(t_c-0)\big] {\bf r}_{ij}^{(c)} \, , \\ 
{\bf p}_j(t_c+0) = {\bf p}_j(t_c-0) + \frac{2}{d^2} \big[ {\bf r}_{ij}^{(c)}\cdot {\bf p}_{ij}(t_c-0)\big] {\bf r}_{ij}^{(c)} \, , \\
\end{array}
\right.
\end{align}
where ${\bf r}_{ij}\equiv {\bf r}_i-{\bf r}_j$ and ${\bf p}_{ij}\equiv \frac{1}{2}\left({\bf p}_i-{\bf p}_j\right)$ are the canonically conjugate positions and momenta of the binary system formed by these two particles.  Hence, the total momentum of this binary system ${\bf p}_i+{\bf p}_j$ and its total energy $\frac{{\bf p}_i^2}{2m} + \frac{{\bf p}_j^2}{2m}$ are conserved during every collision.  The relative position ${\bf r}_{ij}^{(c)}$ is always taken as the one satisfying the minimum image convention.

According to equation~(\ref{eq:elastic_coll}), the momenta of the colliding particles undergo the jumps $\Delta{\bf p}_{ij}^{(c)}={\bf p}_{i}(t_c+0)-{\bf p}_{i}(t_c-0)=-{\bf p}_{j}(t_c+0)+{\bf p}_{j}(t_c-0)$, satisfying Newton's third law (i.e., the principle of action-reaction) because $\Delta{\bf p}_{ij}^{(c)}=-\Delta{\bf p}_{ji}^{(c)}$.  Therefore,
the force exerted on the $i^{\rm th}$ particle by the collisional dynamics can be written in the following form,
\begin{align}
{\bf F}_{ij}(t) = \sum_{c} \Delta{\bf p}_{ij}^{(c)} \, \delta(t-t_c) \, ,
\label{eq:ForceHS}
\end{align}
where the sum extends over all the elastic collisions $\{c\}$.  We note that every collision $c$ involves a specific pair $i\ne j$ of particles.

The dynamics of the hard-sphere system can be numerically simulated using an event-driven algorithm \cite{H97}.  According to this algorithm, the time evolution is driven by the successive elastic collisions undergone by the particles.  At every collision, the two particles involved in the collision change their momenta according to equation~(\ref{eq:elastic_coll}) and they thus start two new free flights~(\ref{eq:free_flight}).  All the times for their possible future collisions with the other particles are computed by solving the quadratic equations $[{\bf r}_i(t)-{\bf r}_k(t)]^2=d^2$ between the position ${\bf r}_i$ of either one or the other of the two spheres that have just undergone a collision, and the positions ${\bf r}_k$ of all the other spheres for $t-t_c>0$.  Among all the times of possible collisions for each sphere, the smallest one is selected for its next collision to happen.  This algorithm is very powerful to simulate the dynamics.

The total energy and the total linear momentum are fixed at the given values $H=E$ and ${\bf P}=0$ by the initial conditions.
At equilibrium, the temperature is thus related to the total energy by $k_{\rm B}T = (2/3)(E/N)$.  

We note that the dynamics of the hard-sphere system is known to be chaotic with sensitivity to initial conditions and temporal disorder respectively characterized by positive Lyapunov exponents and a positive Kolmogorov-Sinai entropy per unit time, which are proportional to the collision frequency \cite{K44,S96,DP97,vBDPD97}.  On this ground, the ergodic and mixing properties have been proved for systems containing two and some higher numbers of hard spheres moving on the torus \cite{SC87,KSS91,KSS92,S04}.

In the numerical simulations, we consider hard spheres of unit diameter $d=1$ and unit mass $m=1$, and the temperature is fixed to the value $k_{\rm B}T=1$.  The numerical results are presented in terms of dimensionless quantities.  The dimensionless particle density is defined as $n_*\equiv nd^3=Nd^3/V$.  The dimensionless positions, momenta, and time are respectively taken as ${\bf r}_{i*}\equiv {\bf r}_i/d$, ${\bf p}_{i*}\equiv{\bf p}_i/\sqrt{mk_{\rm B}T}$, and $t_*\equiv (t/d)\sqrt{k_{\rm B}T/m}$.

\subsection{Equilibrium properties}

The equilibrium statistical distribution of the hard-sphere system is taken as the microcanonical probability distribution~(\ref{eq:microcan}) with ${\bf P}=0$ and a fixed value for $E$.

For a total mass $M=mN$, the equilibrium mass density is equal to $\rho=mn$ with the mean particle density $n=N/V$.

The equation of state for the specific internal energy is given by equation~(\ref{eq:int_energy}) with the binary energy potential~(\ref{eq:u(2)-HS}).  Since the temperature is measured according to equation~(\ref{eq:temperature}), the internal energy per unit mass is given by $e(n,T)=3k_{\rm B}T/(2m)$ and the specific heat capacity at constant specific volume by $c_v=(\partial e/\partial T)_v = 3k_{\rm B}/(2m)$.

The hydrostatic pressure can be computed with the equation of state~(\ref{eq:pressure}) for the force given by equation~(\ref{eq:ForceHS}) as
\begin{align}
\label{eq:pressureHS}
p = nk_{\rm B} T + \lim_{t\to\infty} \frac{1}{3Vt} \sum_{c\,\in\,[0,t]}  {\bf r}_{ij}^{(c)} \cdot \Delta{\bf p}_{ij}^{(c)}\, ,
\end{align}
where the sum extends over all the collisions $c$ occurring in the time interval $0<t_c<t$.  
We note that, for the hard-sphere system, the equation of state for the pressure can be written in the following form,
\begin{align}\label{eq:EoS_pressureHS}
p(n,T) = k_{\rm B} T \, f(n) 
\qquad\mbox{with} \qquad
\left(\frac{\partial f}{\partial T}\right)_n=0 \, .
\end{align}
Since the function $f(n)$ does not depend on the temperature, its dependence on the particle density $n$ provides the complete knowledge of the pressure for this system.  We note that the phase is fluid for the particle densities $0\leq n_*\leq 0.938\pm 0.003$ and crystalline for $1.037\pm 0.003 \leq n_* < \sqrt{2}$.  At fluid-crystal coexistence, the pressure has the dimensionless value $p_{*{\rm FC}}=11.55 \pm 0.11$ \cite{S97}.

We compute the pressure as a function of the density $n_*$ using equation~(\ref{eq:pressureHS}) for a fluid of $N=500$ hard spheres. Statistics are carried out over $10^4$ trajectories with $400$ time steps $\Delta t_*=0.1$ for $n_* = 0.1$ and $0.2$;  $200$ time steps $\Delta t_*=0.1$ for $n_* = 0.3$, $0.4$, and $ 0.5$; and $200$ time steps $\Delta t_*=0.05$ for the other densities. The results for the pressure are given in table~\ref{Tab:P} in comparison with two approximations for the function $f(n)$: first, the truncated virial expansion
\begin{align}
\label{eq:Virial_eos}
f_{\rm V}(n) = \frac{3}{2\pi d^3} \left(x+ x^2+b_3x^3+b_4x^4+b_5x^5+b_6x^6\right)
\qquad\mbox{with}\qquad
x=\frac{2\pi}{3}\, d^3 n \, ,
\end{align}
where $b_3=5/8$, $b_4=18.36/64$, $b_5=28.26/256$, and $b_6=39.53/1024$ \cite{B75}; and secondly, the following Pad\'e approximant
\begin{align}
\label{eq:Padé_eos}
f_{\rm P}(n) = \frac{3x}{2\pi d^3}\left(1 + \frac{a_1x+a_2x^2+a_3x^3}{1+c_1x+c_2x^2}\right)
\end{align}
with the parameters $a_1=1$, $a_2 = 0.076 014$, $a_3 = 0.019 480$, $c_1=-0.548 986$, and $c_2=0.075 647$ \cite{S97}.  The comparison in table~\ref{Tab:P} shows that the second approximation is better than the first one, which is confirmed by the plot of the pressure versus the density in the top left panel of figure~\ref{Fig:TP}.  The truncated virial expansion is a good approximation at small densities, but not at larger densities due to the truncation.  The Pad\'e approximant has been proposed as a way to estimate and sum the higher terms in the virial expansion. For this reason the Pad\'e approximant is a good approximation on the whole range of densities.  Moreover, the fit of the Pad\'e approximant~(\ref{eq:Padé_eos}) to the numerical data from equation~(\ref{eq:pressureHS}) gives the following values for the parameters: $a_1= 0.999891\pm0.002$, $a_2=0.078580\pm 0.007$, $a_3=0.020283\pm 0.009$, $c_1 = -0.547639\pm 0.01$, and $c_2=0.075264\pm 0.004$, which are in excellent agreement with the aforementioned values of Ref.~\cite{S97}.

For the equation of state~(\ref{eq:EoS_pressureHS}), the specific entropy $s$ such that ${\mathrm d}s=({\mathrm d}e+p\, {\mathrm d}v)/T$, the adiabatic speed of sound $c_{ s}$, the specific heat capacity at constant pressure $c_p$, the specific heat ratio $\gamma\equiv c_p/c_v$, and the isothermal compressibility $\chi_T$ are given as follows in terms of the function $f(n)$ and its derivative $f'(n)=({\mathrm d}/{\mathrm d}n)f(n)$ \cite{R99}:
\begin{align}
s &= \frac{k_{\rm B}}{m} \left[ \frac{3}{2} \, \ln e - \int \frac{f(n)}{n^2} \, {\mathrm d}n + \mbox{constant} \right] , \label{eq:entropyHS}\\
c_p&=\frac{3 k_{\rm B}}{2m} \left[1+\frac{2}{3}\frac{f(n)^2}{n^2f'(n)}\right] ,\label{eq:c_P}\\
\chi_T &=	\frac{1}{n k_{\rm B} T f'(n)}\, ,\label{eq:chi_T}\\
c_{s}^2 &= \frac{k_{\rm B}T}{m} \left[ f'(n) + \frac{2}{3} \frac{f(n)^2}{n^2}\right] , \label{eq:c_s}\\
\gamma&=1+\frac{2}{3}\frac{f(n)^2}{n^2f'(n)}=\frac{mc_s^2}{k_{\rm B}T f'(n)}\, . \label{eq:gamma}
\end{align}
Other thermodynamic quantities can be deduced from equations~\eqref{eq:chi_T}-\eqref{eq:gamma} such as the adiabatic compressibility $\chi_s=\chi_T/\gamma$ and the isothermal speed of sound $c_T^2=1/(mn\chi_T)$.

The values of these thermodynamical properties computed from the pressure~(\ref{eq:pressureHS}) are given in table~\ref{Tab:TP} and shown in figure~\ref{Fig:TP}. The results are in agreement with the literature~\cite{AA83,R99}.

Furthermore, we have also computed the static structure factor~(\ref{eq:ssf}) in simulations using $N=500$ hard spheres at various densities with the same parameter values as mentioned here above and for the wave numbers $q=2\pi(n_x^2+n_y^2+n_z^2)^{1/2}/L$ with the integer values $n_x,n_y,n_z\in(1,2,...,25)$.  The discrete values of $q$ are a consequence of the choice of periodic boundary conditions on the dynamics and the observables. The results for the static structure factor are shown in figure~\ref{Fig:SSF} and they are in agreement with those of Ref.~\cite{AAY83} and with the Percus-Yevick approximation~\cite{W63,H09}.

The inset of figure~\ref{Fig:SSF} depicts the small-$q$ limit of $S(q)$.  An extrapolation of $S(q)$ at $q=0$ is obtained with a linear regression over the $q^2$ dependence of the values and the isothermal compressibility $\chi_T$ is evaluated using equation~\eqref{eq:isocompSq}.  These values of $\lim_{q\rightarrow 0}S(q)$ and $\chi_T$ are reported in table~\ref{Tab:SSF}, which shows that there is very good agreement with the corresponding values obtained from the Percus-Yevick approximation: 
\begin{align}
\chi_T = \frac{1}{n  k_{\rm B} T}\frac{(1- y)^4}{(1+2 y)^2} \, ,
\qquad\mbox{where}\qquad
y=\frac{x}{4}=\frac{\pi}{6}\, d^3 n
\end{align}
is the packing fraction. Furthermore, the values of table~\ref{Tab:SSF} for the isothermal compressibility $\chi_T$ obtained from the static structure factor $S(q)$ are in good agreement with those of table~\ref{Tab:TP} for $\chi_T$ computed with equation~(\ref{eq:chi_T}) and the pressure~(\ref{eq:pressureHS}).

\subsection{Nonequilibrium properties}

The transport coefficients can be computed using the Einstein-Helfand formulas~(\ref{eq:etaH})-(\ref{eq:longviscH}) with the Helfand moments~(\ref{eq:Helfand}) corresponding to the global currents~(\ref{eq:Jab})-(\ref{eq:Jaq}).  For the hard-sphere system, these Helfand moments can be calculated for particles following the free flights~(\ref{eq:free_flight}) and interacting by the forces~(\ref{eq:ForceHS}).  The Helfand moments associated with momentum and energy transport are thus respectively given by
\begin{align}
\Delta{\mathbb G}^{ab}(t)&= \sum_{(c\to c+1)\,\in\,[0,t]} \sum_i\frac{p_i^a(t_c+0)}{m}\, p_i^b(t_c+0) \, (t_{c+1}-t_c) + \sum_{c\, \in \, [0,t]}r_{ij}^{a (c)} \, \Delta {p}_{ij}^{b (c)} \,,\label{eq:DGabHS}\\
\Delta{\mathbb G}_{e}^a(t)&=\sum_{(c\to c+1)\,\in\,[0,t]} \sum_{i} \frac{{p}^a_i(t_c+0) }{m}\, \frac{[{\bf p}_{i}(t_c+0)]^2}{2m}\, (t_{c+1}-t_c) + \sum_{c\, \in \, [0,t]}  r^{a (c)}_{ij} \, \Delta{p}_{ij}^{b (c)} \, \frac{{p}^{b (c)}_{i}+{p}^{b (c)}_{j}}{2m}\, ,\label{eq:DGeaHS}
\end{align}
where the first sum is carried out over the free flights during the time interval $[0,t]$ (or their piece if the free flight includes the initial time $t=0$ or the final time $t$) and the second sum is over the collisions occurring for $0\leq t_c \leq t$.

Using these Helfand moments in molecular dynamics simulations, we compute the transport coefficients in the same code as the pressure.  Their values are reported in table~\ref{Tab:TC} and shown in figure~\ref{Fig:TC} for a fluid of $N=500$ hard spheres with  statistics carried out over $10^4$ trajectories with $400$ time steps $\Delta t_*=0.1$ for $n_* = 0.1$ and $0.2$;  $200$ time steps $\Delta t_*=0.1$ for $n_* = 0.3$, $0.4$, and $ 0.5$; and $200$ time steps $\Delta t_*=0.05$ for the other densities.  The variance of the Helfand moments is computed during the time evolution with the Welford algorithm~\cite{W62} and the slope of the variance as a function of time is obtained using a linear least square regression to evaluate the corresponding transport coefficient. The error is estimated as the difference of their values obtained from a regression over the second half of the time interval.  We note that $N=500$ is a large enough particle number for the computed quantities to be closer to their large-system limit than statistical errors on time averages.  The numerical values are compared in table~\ref{Tab:TC} and figure~\ref{Fig:TC} with the predictions of Enskog theory, which are given in appendix~\ref{app:Enskog} for each transport coefficient.  The results for the transport coefficients are in agreement with the literature~\cite{AGW70,SH03}.

Now, the issue is to test numerically these values obtained using the Einstein-Helfand formulas with the locations of the poles of the hydrodynamic spectral functions at the complex frequencies (\ref{eq:heat_pole})-(\ref{eq:shear_pole}).  For this purpose, we perform molecular dynamics simulations to compute the intermediate scattering function~(\ref{eq:isf}) and the longitudinal and transverse components~(\ref{eq:vcf+vsf-l+t}) of the time-dependent correlation functions~(\ref{eq:vcf}) characterizing momentum fluctuations.  By taking their temporal Fourier transform, we obtain the corresponding dynamic structure factor~(\ref{eq:dsf}) and the longitudinal and transverse components of the spectral functions~(\ref{eq:vsf}) for momentum fluctuations, respectively.  The simulation parameters are the following: For $n_*=0.144$, we use $N=2048$ hard spheres and $10^3$ trajectories with $500$ time steps $\Delta t_*=0.1$; for $n_*=0.476$, $N=500$ hard spheres and $10^4$ trajectories with $400$ time steps $\Delta t_*= 0.05$; and for $n_*=0.884$, $N=500$ hard spheres and $10^4$ trajectories with $150$ time steps $\Delta t_*=0.02$. We have computed the correlation functions for the smallest values of $q=2\pi(n_x^2+n_y^2+n_z^2)^{1/2}/L$. 

The correlation and spectral functions are shown in figures~\ref{Fig:CC0144},~\ref{Fig:CC0471}, and~\ref{Fig:CC0884} for the densities $n_*=0.144$, $0.476$, and $0.884$, respectively.  The Rayleigh and Brillouin resonances appear in the dynamic structure factor.  The Rayleigh resonance at $\omega=0$ is suppressed in the longitudinal component of the spectral function for momentum fluctuations because of its relation~(\ref{eq:lca_hydro}) to the dynamic structure factor.  The transverse component of the spectral function for momentum fluctuations presents a resonance around zero frequency associated with the diffusive mode of shear viscosity.  We observe that the width of each resonance indeed behaves as the corresponding diffusivity.
Numerical integrations show that the sum rules given by equations~\eqref{eq:sr_dsf} and~\eqref{eq:sr_vsf-l+t} are satisfied. The results from the numerical simulations are compared with the theoretical predictions of equations~\eqref{eq:dsf_hydro},~\eqref{eq:lca_hydro}, and~\eqref{eq:tvsf_hydro} with the parameter values provided by the formula~(\ref{eq:pressureHS}) for the pressure and the Einstein-Helfand formulas~(\ref{eq:etaH})-(\ref{eq:longviscH}) with the Helfand moments~(\ref{eq:DGabHS})-(\ref{eq:DGeaHS}) for the transport coefficients.  The results are in agreement with the literature~\cite{BY80,AAY83}.

To investigate the hydrodynamic limit of vanishing $q$, we consider the intermediate scattering function $F(q,t)$ for the smallest $q$, given by $q_{\rm min}\equiv2\pi(n/N)^{1/3}$ for values of $N=500, 1372, 2916$, and  $5324$ at $n_*=0.144$. The results are shown in figure~\ref{Fig:HL}, where we observe that the intermediate scattering function obtained with molecular dynamics is the closest to the hydrodynamic prediction (dotted line) for the smallest value of the wave number $q$, i.e., for the largest possible wave length, as expected.

The poles of the dynamic structure factor and the longitudinal and transverse components of the spectral functions for momentum fluctuations are obtained by fitting the rational functions~(\ref{eq:rational_fn}) to these functions.  The degrees of the polynomials at the denominator and numerator are taken as $6\leq M\leq 24$ and $1\leq N \leq M-2$ for the dynamic structure factor and the longitudinal component of the spectral functions for momentum fluctuations and $2\leq M\leq 10$ and $0\leq N \leq M-2$ for the transverse component of the spectral functions for momentum fluctuations.  For each function, the coefficients $\{a_k\}$ and $\{b_k\}$ are fitted with a nonlinear least square method.  For any choice of $M$ and $N$, the poles and their complex conjugates expected from the dispersion relations (\ref{eq:heat_pole})-(\ref{eq:shear_pole}) of the hydrodynamic modes are found in the physically relevant region of the plane of complex frequencies.  In this way, the poles underlying the Rayleigh and Brillouin resonances can be identified as corresponding to the frequencies (\ref{eq:heat_pole})-(\ref{eq:sound_poles}) of the heat and sound modes, and the poles for the resonance observed in the transverse component of the spectral function for momentum fluctuations to the frequency (\ref{eq:shear_pole}) of the shear mode.  The real part of the poles from the central resonances is nearly equal to zero, while those of the Brillouin doublet is not.  The location of the poles is determined by averaging over the poles obtained for different degrees $M$ and $N$.  The error on the location is given by the standard error on the mean value, namely, the standard deviation divided by the square root of the number of fitted poles.  The determination of the poles is repeated for different values of the wave number $q$ for a fluid of $N=2048$ hard spheres at density $n_*=0.144$, and $N=500$ hard spheres at densities $n_*=0.476$ and $n_*=0.884$.  Equations~(\ref{eq:q-coefficients}) give the $q$-dependent speed of sound, acoustic attenuation coefficient, thermal diffusivity, and transverse kinematic viscosity, from which the shear viscosity is obtained as $\eta=\rho\nu$. These values are given in tables~\ref{Tab:CC0144},~\ref{Tab:CC0471}, and~\ref{Tab:CC0884}.  The extrapolation of these values to $q=0$ are obtained with a least square linear regression over the dependence on $q^2$ for the quantities of interest, as plotted in figures~\ref{Fig:LR0144},~\ref{Fig:LR0471}, and~\ref{Fig:LR0884}.  Finally, the values obtained in this way from the spectral functions are compared with those given by the equation of state for the pressure and the Einstein-Helfand formulas for the transport coefficients shown in the last line of tables~\ref{Tab:CC0144}-\ref{Tab:CC0884} and the open circle in figures~\ref{Fig:LR0144}-\ref{Fig:LR0884}.  There is a good agreement between these values.

\section{Conclusion and perspectives}
\label{sec:conclusion}

In this paper, we have developed and compared two approaches to calculate and numerically compute the transport coefficients in bulk phases of matter.  

On the one hand, the local-equilibrium approach allows us to derive the macroscopic equations of hydrodynamics, including the transport coefficients given by Einstein-Helfand formulas.  With these formulas, the transport coefficients can be computed as the diffusivities of Helfand moments associated with the transported quantities.

On the other hand, we have used another approach providing the damping rates of the time-dependent correlation functions for the equilibrium hydrodynamic fluctuations.  These damping rates are computed by analytic continuation of the corresponding spectral functions from real to complex frequencies in order to identify the poles of the spectral functions.  These poles determine the resonances generating relaxation towards equilibrium in matter and their imaginary part gives the width of the resonances and, thus, the damping rates, which depend on the transport coefficients.  Taking functional derivatives with respect to initial conditions applied to the hydrodynamic equations obtained in the local-equilibrium approach, we have shown that the time-dependent correlation functions are ruled by the linearized hydrodynamic equations around equilibrium in accord with Onsager's hypothesis of regression of fluctuations.  As a consequence, the complex frequencies of the poles of the spectral functions should precisely correspond to the dispersion relations of the hydrodynamic modes in the bulk phase of interest, these dispersion relations being expressed in terms of the transport coefficients.   In this way, the computation of the spectral functions and their poles provide a general method to test numerically the values of the transport coefficients obtained by the Einstein-Helfand formulas.

In the case of fluids, there are five hydrodynamic modes: the heat mode and the two sound modes, giving the Rayleigh and the two Brillouin resonances in the dynamic structure factor, i.e., the spectral function characterizing the fluctuations of density; and the two shear modes giving a sole resonance in the transverse component of the spectral function for the fluctuations of momentum.  

We have systematically carried out this comparison between the two approaches in the case of the hard-sphere fluid.  The molecular dynamics simulation of this system has been performed using an event-driven algorithm.  First, we have computed the pressure and therefrom derived equilibrium properties since the dispersion relations of the sound modes depend not only on the transport coefficient, but also on the speed of sound and the heat capacities.  We have also computed the shear and bulk viscosities and the heat conductivity as a function of the particle density, using the Einstein-Helfand formulas.

Next, we have numerically calculated the time-dependent correlation functions for the equilibrium fluctuations of the particle density (i.e., the intermediate scattering function) and of the transverse momentum density.  By numerical Fourier transform in time, we have obtained the corresponding spectral functions.  These functions have been computed for three values of the mean particle density in the fluid phase.  For each density, the procedure has been repeated for several values of the wave number taken as small as possible to reach the hydrodynamic regime.  In each case, the poles at complex frequencies have been located by fitting rational functions to the spectral functions.  In this way, we have obtained numerical evaluations for the dispersion relations of the hydrodynamic modes as a function of the wave number.  Their extrapolations to zero wave number are in good agreement with the prediction of hydrodynamics based on the transport coefficients obtained by the Einstein-Helfand formulas in the local-equilibrium approach.  

This agreement shows the consistency of the two approaches based on local equilibrium and on spectral functions in the hydrodynamic limit.  This numerical test also shows that the agreement is obtained in the limit of small enough values for the wave number $q$, possibly requiring to increase the number $N$ of hard spheres used in molecular dynamics simulation up to several thousands, which remains very small compared to the Avogadro number.  Therefore, our results provide numerical support to the methods developed in the two approaches to transport processes.

We note that the identification of the poles corresponding to the hydrodynamic resonances is a method similar to the one used to obtain the Pollicott-Ruelle resonances in models of deterministic diffusion such as the Lorentz gas and related systems \cite{G96,G98,GCGD01,G22}.  In this analogy, the hydrodynamic modes can be considered as generalized eigenmodes for the microscopic Liouvillian dynamics of the $N$-particle system.

Furthermore, the approaches developed in the present work to calculate the transport coefficients can also be applied to other bulk phases of matter than the fluid phase and, in particular, to the crystalline phase, where there exist eight hydrodynamic modes and a larger set of transport coefficients because of broken continuous symmetries in these anisotropic phases of matter \cite{MG20,MG21}.  In future work, we hope to report on these perspectives opened by the present study.


\section*{Acknowledgements}

The authors thank the Universit\'e Libre de Bruxelles (ULB) and the Fonds de la Recherche Scientifique de Belgique (F.R.S.-FNRS) for support in this research. J.~M. is a Postdoctoral Researcher of the Fonds de la Recherche Scientifique de Belgique (F.R.S.-FNRS).  Computational resources have been provided by the Consortium des Equipements de Calcul Intensif (CECI), funded by the Fonds de la Recherche Scientifique de Belgique (F.R.S.-FNRS) under Grant No. 2.5020.11 and by the Walloon Region.


\appendix

\section{Regression of fluctuations around equilibrium}
\label{app:rh-leq}

Around thermodynamic equilibrium, the hypothesis of regression of fluctuations can be justified using the local-equilibrium approach summarized in section~\ref{sec:leq}.  Within this approach, it is established that the nonequilibrium mean values~(\ref{eq:neq-mean-value}) of the microscopic densities obey the macroscopic equations of hydrodynamics on large spatiotemporal scales \cite{M58,McL63,R66,P68,Z74,AP81,OL79,S14,DLW20,MG20,MG21,MG23,G22}.  During the stage of relaxation towards equilibrium, these macroscopic equations can be linearized around the state of equilibrium, where the fluid has zero velocity and uniform profiles for density and temperature.

Now, in the local-equilibrium approach, the initial condition of the nonequilibrium probability distribution~(\ref{eq:leq-prob-t}) is given by the local-equilibrium distribution~(\ref{eq:leq-prob}) with initial values $\lambda^\alpha_0({\bf r})=\lambda^\alpha({\bf r},t=0)$ for the conjugate fields.  Therefore, the time-dependent correlation functions of the microscopic densities can be obtained by taking the functional derivatives of the nonequilibrium mean values~(\ref{eq:neq-mean-value}) with respect to the initial values of the conjugate fields as
\begin{align}
\langle\delta\hat c^\alpha({\bf r},t) \, \delta\hat c^\beta({\bf r}^\prime,0)\rangle_{{\rm leq},\boldsymbol{\lambda}_0} = - \frac{\delta\langle\hat c^\alpha({\bf r})\rangle_t}{\delta\lambda^\beta({\bf r}^{\prime},0)} \, ,
\end{align}
where $\delta\hat c^\alpha \equiv \hat c^\alpha- \langle \hat c^\alpha\rangle_{{\rm leq},\boldsymbol{\lambda}_0}$.
Because of the mean local conservation equations~(\ref{eq:av-local-eqs}), the correlation functions should satisfy the following equations,
\begin{align}
\partial_t \, \langle\delta\hat c^\alpha({\bf r},t) \, \delta\hat c^\beta({\bf r}^\prime,0)\rangle_{{\rm leq},\boldsymbol{\lambda}_0}= \nabla^a \, \frac{\delta\langle\hat J_{c^\alpha}^a({\bf r})\rangle_t}{\delta\lambda^\beta({\bf r}^{\prime},0)} \, .
\label{eq:cf-eqs}
\end{align}
However, the mean values over the nonequilibrium probability distribution~(\ref{eq:leq-prob-t}) can also be expressed in terms of the conjugate fields $\lambda^\alpha({\bf r},t)$ at time $t$, so that
\begin{align}
\frac{\delta\langle\hat J_{c^\alpha}^a({\bf r})\rangle_t}{\delta\lambda^\beta({\bf r}^{\prime},0)} = \int \frac{\delta\langle\hat J_{c^\alpha}^a({\bf r})\rangle_t}{\delta\lambda^\gamma({\bf r}^{\prime\prime},t)} \, \frac{\delta\lambda^\gamma({\bf r}^{\prime\prime},t)}{\delta\lambda^\beta({\bf r}^{\prime},0)} \, {\mathrm d}{\bf r}^{\prime\prime}
\label{eq:J-lambda(t)}
\end{align}
and
\begin{align}
\label{eq:ct-c0}
\int \langle\delta\hat c^\alpha({\bf r},t) \, \delta\hat c^\beta({\bf r}^\prime,0)\rangle_{{\rm leq},\boldsymbol{\lambda}_0} \, \delta\lambda^\beta({\bf r}^{\prime},0) \, {\mathrm d}{\bf r}^{\prime} = \int \langle\delta\hat c^\alpha({\bf r},0) \, \delta\hat c^\beta({\bf r}^\prime,0)\rangle_{{\rm leq},\boldsymbol{\lambda}_t} \, \delta\lambda^\beta({\bf r}^{\prime},t) \, {\mathrm d}{\bf r}^{\prime} \, .
\end{align}
In equation~(\ref{eq:ct-c0}), the left-hand side contains the time-dependent correlation functions and the variation of the conjugate fields at time $t=0$, but the right-hand side is expressed in terms of the equal-time correlation functions between the two observables, and the variation of the conjugate fields at time $t$.  The important point is that the equal-time correlation functions  characterize the spatial correlations at given time around local equilibrium.  The equilibrium expressions of these correlations are known at equilibrium \cite{LL80b,OS06}, where they behave as $\langle\delta\hat c^\alpha({\bf r},0) \, \delta\hat c^\beta({\bf r}^\prime,0)\rangle_{\rm eq} \simeq \Gamma_{\rm eq}^{\alpha\beta} \, \delta({\bf r}-{\bf r}^{\prime})$ over scales larger than the characteristic spatial correlation length of the fluid.

In order to obtain the time-dependent correlation functions around equilibrium, we may consider the limit where $\boldsymbol{\lambda}_0=\boldsymbol{\lambda}_t=\boldsymbol{\lambda}_{\rm eq}$ in equation~(\ref{eq:ct-c0}) to get
\begin{align}
\frac{\delta\lambda^\alpha({\bf r},t)}{\delta\lambda^\beta({\bf r}^{\prime},0)}\Bigg\vert_{\rm eq} \simeq \left(\boldsymbol{\Gamma}_{\rm eq}^{-1}\right)^{\alpha\gamma} \langle\delta\hat c^\gamma({\bf r},t) \, \delta\hat c^\beta({\bf r}^\prime,0)\rangle_{\rm eq} \, ,
\label{eq:lambda(t)-lambda(0)}
\end{align}
where $\left(\boldsymbol{\Gamma}_{\rm eq}^{-1}\right)^{\alpha\gamma}\Gamma_{\rm eq}^{\gamma\beta} = \delta^{\alpha\beta}$.  Replacing equation~(\ref{eq:J-lambda(t)}) into equation~(\ref{eq:cf-eqs}) and using the result~(\ref{eq:lambda(t)-lambda(0)}) in the limit $\boldsymbol{\lambda}_0=\boldsymbol{\lambda}_t=\boldsymbol{\lambda}_{\rm eq}$, we obtain a closed set of linear equations for the time-dependent correlation functions. These equations are nothing else than the linearized hydrodynamic equations over large spatiotemporal scales.  In this way, the hypothesis of regression of fluctuations can be justified around equilibrium.

We note that the equilibrium spatial fluctuations of the fields $\delta\hat c^{\alpha}=(\delta\hat \rho,\delta\hat\epsilon,\delta\hat g^a)$ are not all statistically independent of each other, so that the matrix $\boldsymbol{\Gamma}_{\rm eq}=(\Gamma_{\rm eq}^{\alpha\beta})$ is not diagonal.  However, new fluctuating fields can be defined by some linear transformation $\delta\hat\phi^\alpha=A^{\alpha\beta} \delta\hat c^\beta$ that diagonalizes the matrix $\boldsymbol{\Gamma}_{\rm eq}$.  Consequently, the new fluctuating fields are statistically independent.  This is in particular the case for $\delta\hat\phi^{\alpha}=(\delta\hat\rho,\delta\hat T,\delta\hat v^a)$, where $\delta\hat T =c_v^{-1}\left[\delta\hat e-(\partial e/\partial\rho)_T \,\delta\hat\rho\right]$ with $\delta\hat e=\rho^{-1}\delta\hat\epsilon$, and $\delta\hat v^a=\rho^{-1}\delta\hat g^a$ \cite{LL80b,OS06}.  At equilibrium, the equal-time autocorrelation functions of these statistically independent fluctuating fields are given on large spatial scales by
\begin{align}
\langle\delta\hat\rho({\bf r},0)\, \delta\hat\rho({\bf r}^{\prime},0)\rangle_{\rm eq} &\simeq \rho^2 k_{\rm B}T \, \chi_T \, \delta({\bf r}-{\bf r}^{\prime}) \, , \\
\langle\delta\hat T({\bf r},0)\, \delta\hat T({\bf r}^{\prime},0)\rangle_{\rm eq} &\simeq \frac{k_{\rm B}T^2}{\rho \, c_v} \, \delta({\bf r}-{\bf r}^{\prime}) \, , \\
\langle\delta\hat v^a({\bf r},0)\, \delta\hat v^b({\bf r}^{\prime},0)\rangle_{\rm eq} &\simeq \frac{k_{\rm B}T}{\rho} \, \delta^{ab} \, \delta({\bf r}-{\bf r}^{\prime}) \, ,
\end{align}
where $\chi_T$ is the isothermal compressibility and $c_v$ the specific heat capacity at constant specific volume \cite{LL80b,OS06}.

\section{Correlation and spectral functions in the hydrodynamic regime}
\label{app:hydro-cf+sf}

As justified in appendix~\ref{app:rh-leq}, the time-dependent correlation functions of the fluctuating fields can be calculated around equilibrium using the hypothesis of regression of fluctuations, according to which these correlation functions obey the linearized equations of macroscopic hydrodynamics.  To simplify the calculation, we consider the statistically independent fluctuating fields $\delta\hat\phi^{\alpha}=(\delta\hat\rho,\delta\hat T,\delta\hat v^a)$.  The linearized equations ruling the dynamics of these fields can be deduced from the macroscopic hydrodynamic equations \cite{F75,BP76,BY80}.

These latter are given by the local conservation equations for mass, energy, and linear momentum \cite{GM84}:
\begin{align}
\partial_t\rho+\nabla^a(\rho\, v^a)&=0 \, ,   \label{dissip-eq-mass}\\
\partial_t\epsilon + \nabla^a\big[(\epsilon + p ) v^a + \, \Pi^{ab} \, v^b +{\cal J}^a_q \big]& = 0 \, , \label{dissip-eq-etot}\\
\partial_t(\rho\, v^a)+\nabla^b\big(\rho\, v^a v^b + p \, \delta^{ab} + \Pi^{ab}\big) &= 0 \, , \label{dissip-eq-g}
\end{align}
where $\epsilon=(1/2) \rho (v^a)^2+ \rho e $ is the sum of the kinetic and internal energy densities, while the viscous pressure tensor and the heat current density can be respectively expressed as
\begin{align}\label{dissip-Pi+J_q}
\Pi^{ab}= -\eta \left( {\nabla}^a{v}^b+\nabla^b{v}^a- \frac{2}{3} \, {\nabla}^c{v}^c\delta^{ab}\right)-\zeta\, {\nabla}^c v^c\, \delta^{ab} 
\qquad\mbox{and}\qquad
{\cal J}^a_{q} =-\kappa\, \nabla^aT 
\end{align}
in terms of the shear viscosity $\eta$, the bulk viscosity $\zeta$, and the heat conductivity $\kappa$.

The nonlinear equations (\ref{dissip-eq-mass})-(\ref{dissip-eq-g}) are linearized around the equilibrium macrostate where the fluid is at rest ($v^a=0$) and has uniform profiles for mass density $\rho$, temperature $T$, pressure $p$, specific internal energy $e$, and specific entropy $s$.  In the following, the symbol $\delta$ will denote the deviation of some field with respect to its equilibrium profile.  Furthermore, we note that the linearized equation~(\ref{dissip-eq-etot}) for the energy density leads, first, to the equation for the specific internal energy by using $\delta\epsilon \simeq e\,\delta\rho + \rho\,\delta e$ since the fluid velocity is equal to zero in the reference macrostate and, next, to the equation for the specific entropy because of the Gibbs thermodynamic relation $\delta e = T\,\delta s + p\, \delta\rho/\rho^2$.  As a consequence, the linearized equations of hydrodynamics to consider are the following ones,
\begin{align}
\partial_t\, \delta\rho &= -\rho\, \nabla^a \delta v^a  \, , \label{lin-dissip-eq-mass}\\
\rho\, T \, \partial_t\, \delta s &= \kappa \, \nabla^2 \delta T  \, ,  \label{lin-dissip-eq-s}\\
\rho\, \partial_t \delta v^a &=-\nabla^a\delta p +\eta \nabla^2 \delta v^a + \Big(\zeta+\frac{1}{3}\, \eta\Big)\nabla^a\nabla^b \delta v^b \, . \label{lin-dissip-eq-v}
\end{align}
The solutions of these linearized equations give the so-called hydrodynamic modes \cite{B75,RD77,R98,G22}.

In order to obtain a closed set of equations for the fields $(\delta\rho,\delta T,\delta v^a)$, the specific entropy $s$ and the hydrostatic pressure $p$ are taken as functions of the density $\rho$ and the temperature $T$, so that
\begin{align}
\delta s &= \left(\frac{\partial s}{\partial \rho}\right)_T \delta \rho+\left(\frac{\partial s}{\partial T}\right)_\rho \delta T 
\qquad\mbox{and}\qquad
\delta p = \left(\frac{\partial p}{\partial \rho}\right)_T \delta \rho+\left(\frac{\partial p}{\partial T}\right)_\rho \delta T \, .
\end{align}
Using standard thermodynamic relations, the coefficients can be expressed in terms of the specific heat capacities $c_p$ and $c_v$, their ratio $\gamma=c_p/c_v$, the isothermal compressibility $\chi_T$, the adiabatic speed of sound $c_s$, and the thermal expansion coefficient $\alpha=-\rho^{-1}(\partial\rho/\partial T)_p$, so that we obtain
\begin{align}\label{eq:ds+dp}
\delta s &= \frac{c_v}{T} \left( -\frac{\gamma-1}{\rho\, \alpha} \, \delta \rho +  \delta T \right)
\qquad\mbox{and}\qquad
\delta p = \frac{c_s^2}{\gamma} \left(\delta \rho+\rho\, \alpha \, \delta T\right) \, .
\end{align}
Accordingly, we find the linear equation for the temperature $\delta T$ as
\begin{align}
\partial_t\, \delta T = - \frac{\gamma-1}{\alpha} \nabla^a\delta v^a + \gamma D_T \, \nabla^2 \delta T    \label{lin-dissip-eq-T}
\end{align}
with the thermal diffusivity $D_T \equiv \kappa/(\rho c_p)$.

The equations~(\ref{lin-dissip-eq-mass}), (\ref{lin-dissip-eq-v}) with $\delta p$ given in~(\ref{eq:ds+dp}), and~(\ref {lin-dissip-eq-T}) form the closed set of equations we need for the fields $(\delta\rho,\delta T,\delta v^a)$.

These linearized equations can be solved by considering the spatial Fourier transform of the fields such that
\begin{align}
f({\bf q},t) = \int f({\bf r},t) \, {\rm e}^{{\mathrm i}{\bf q}\cdot{\bf r}}  \, {\mathrm d}{\bf r} 
\qquad\mbox{and}\qquad
f({\bf r},t) = \frac{1}{(2\pi)^3} \int f({\bf q},t) \, {\rm e}^{-{\mathrm i}{\bf q}\cdot{\bf r}}  \, {\mathrm d}{\bf q} \, .
\end{align}
We note that the spatial Fourier transform of the velocity field can be decomposed into its longitudinal and transverse components as
\begin{align}
v^a = \frac{q^a}{q} \, v_{\rm l} + v_{\rm t}^a 
\qquad\mbox{with}\qquad
v_{\rm t}^a = \left( \delta^{ab} - \frac{q^a q^b}{q^2}\right) v^b \, .
\end{align}
Accordingly, the spatial Fourier transform of the linearized equations for the fluid velocity can be decomposed into the following  three equations:
\begin{align}
\partial_t \, \delta v_{\rm l} &= {\mathrm i}\frac{c_s^2}{\rho\, \gamma} q \, \delta\rho + {\mathrm i}\frac{\alpha c_s^2}{\rho} q \, \delta T - D_v q^2 \, \delta v_{\rm l} \, , \\
\partial_t \, \delta v_{\rm t}^a &= - \nu q^2 \, \delta v_{\rm t}^a  \, ,
\end{align}
where $D_v\equiv(\zeta+\frac{4}{3}\eta)/\rho$ and $\nu\equiv\eta/\rho$ are respectively the longitudinal and transverse kinematic viscosities; and $a=1,2$.  Consequently, the two transverse components of the fluid velocity ($\delta v_{\rm t}^a$ with $a=1,2$) are decoupled from its longitudinal component $\delta v_{\rm l}$ and, thus, from the three fields $(\delta\rho,\delta T, \delta v_{\rm l})$, which remain coupled together.

Next, we carry out a Laplace transform in time such that
\begin{align}
\tilde f({\bf q},z) = \int_0^{\infty} f({\bf q},t) \, {\rm e}^{-zt} \, {\mathrm d}t
\qquad\mbox{and}\qquad
f({\bf q},t) = \frac{1}{2\pi{\mathrm i}} \int_{c-{\mathrm i}\infty}^{c+{\mathrm i}\infty} \tilde f({\bf q},z) \, {\rm e}^{zt} \, {\mathrm d}z \, ,
\end{align}
where $c$ is a constant larger than the real part ${\rm Re}\, z_r$ of all the singularities of the function $\tilde f({\bf q},z)$. The equations for the two transverse components of the fluid velocity become
\begin{align}\label{eq:lin-vt-qz}
\left( z + \nu q^2 \right) \delta\tilde v_{\rm t}^a ({\bf q},z) = \delta v_{\rm t}^a ({\bf q},0)
\end{align}
 (for $a=1,2$) and the three other equations can be cast into the following matrix form,
\begin{align}\label{eq:lin-rho-T-vl-qz}
\left[
\begin{array}{lll}
z & -{\mathrm i}\rho q & 0 \\
0 & z+\gamma D_T q^2 & -{\mathrm i} \frac{\gamma-1}{\alpha} q \\
-{\mathrm i} \frac{c_s^2}{\rho\gamma} q & -{\mathrm i} \frac{\alpha c_s^2}{\gamma} q & z+D_v q^2 \\
\end{array}
\right]
\left[
\begin{array}{l}
\delta\tilde\rho({\bf q},z) \\
\delta\tilde T({\bf q},z) \\
\delta\tilde v_{\rm l}({\bf q},z) \\
\end{array}
\right]
=
\left[
\begin{array}{l}
\delta\rho({\bf q},0) \\
\delta T({\bf q},0) \\
\delta v_{\rm l}({\bf q},0) \\
\end{array}
\right] .
\end{align}

Introducing the five-dimensional column matrix $\delta\boldsymbol{\phi}=(\delta\rho,\delta T,\delta v_{\rm l},\delta v_{\rm t}^1,\delta v_{\rm t}^2)^{\rm T}$, where $^{\rm T}$ denotes the transpose and similarly for $\delta{\boldsymbol{\tilde\phi}}$, equations~(\ref{eq:lin-vt-qz}) and~(\ref{eq:lin-rho-T-vl-qz}) take the following matrix form,
\begin{align}\label{eq:M-phi=phi}
\boldsymbol{\mathsf M}(q,z) \cdot \delta{\boldsymbol{\tilde\phi}}({\bf q},z)= \delta\boldsymbol{\phi}({\bf q},0) \, ,
\end{align}
which can be solved by matrix inversion.

Here, the deviations $\delta\boldsymbol{\phi}$ of the fields can be replaced with the fluctuating fields $\delta\boldsymbol{\hat\phi}$ by using the hypothesis of regression of fluctuations.  After multiplying the so-modified equation~(\ref{eq:M-phi=phi}) on its left-hand side by the inverse of the matrix $\boldsymbol{\mathsf M}^{-1}$ and on its right-hand side by $\delta\boldsymbol{\hat\phi}^{\dagger}({\bf q},0)=\delta\boldsymbol{\hat\phi}^{{\rm T}*}({\bf q},0)$, and taking the statistical average $\langle\cdot\rangle_{\rm eq}$ with respect to the equilibrium probability distribution, we find that
\begin{align}
\langle \delta{\boldsymbol{\hat{\tilde\phi}}}({\bf q},z)\, \delta\boldsymbol{\hat\phi}^{\dagger}({\bf q},0)\rangle_{\rm eq} = \boldsymbol{\mathsf M}^{-1}(q,z) \cdot 
\langle \delta{\boldsymbol{\hat\phi}}({\bf q},0)\, \delta\boldsymbol{\hat\phi}^{\dagger}({\bf q},0)\rangle_{\rm eq} \, .
\end{align}
Since the equal-time correlation matrix on the right-hand side is diagonal, there is no coupling between the matrix elements of $\boldsymbol{\mathsf M}^{-1}$.  Considering the density fluctuations, we find that the Laplace transform $\tilde F(q,z)$ of the intermediate scattering function~(\ref{eq:isf}) is related to the static structure factor~(\ref{eq:ssf}) by the diagonal element associated with density in the inverse matrix:
\begin{align}\label{eq:tilde F(q,z)}
\tilde F(q,z) = [\boldsymbol{\mathsf M}^{-1}(q,z) ]_{\rho\rho} \, S(q) \, .
\end{align}
In the complex plane of the variable $z$, this function has poles corresponding to the zeros of the determinant of the matrix: $\det \boldsymbol{\mathsf M}(q,z)=0$.  They are located at the values $z_r={\mathrm i}\omega_r$ corresponding to the complex frequencies (\ref{eq:heat_pole})-(\ref{eq:shear_pole}).  With the decoupling of the two transverse components of velocity, the matrix $\boldsymbol{\mathsf M}(q,z)$ can be reduced to the $3\times 3$ matrix in equation~(\ref{eq:lin-rho-T-vl-qz}).  Its determinant can be factorized as $[z-{\mathrm i}\omega_0(q)][z-{\mathrm i}\omega_+(q)][z-{\mathrm i}\omega_-(q)]$ in terms of the dispersion relations (\ref{eq:heat_pole}) and (\ref{eq:sound_poles}) for the heat mode and the two sound modes.  Hence, inverting this $3\times 3$ matrix leads to the following expression \cite{BP76,BY80}:
\begin{align}\label{eq:tilde F(q,z)-bis}
\tilde F(q,z) = \frac{(z+\gamma D_Tq^2)(z+D_vq^2)+(1-\gamma^{-1})(c_sq)^2}{[z-{\mathrm i}\omega_0(q)][z-{\mathrm i}\omega_+(q)][z-{\mathrm i}\omega_-(q)]} \, S(q) \, .
\end{align}

Now, the intermediate scattering function $F(q,t)$ can be obtained by taking the inverse Laplace transform of the function~(\ref{eq:tilde F(q,z)-bis}).  Expanding the residues of the function in powers of the wave number $q$ and keeping the leading terms up to first order in $q$, the following time dependence is obtained for the intermediate scattering function \cite{BP76}:
\begin{align}
\frac{F(q,t)}{S(q)}&=\left(1-\frac{1}{\gamma}\right) {\rm e}^{-D_T q^2 |t|}+\frac{1}{\gamma}\, {\rm e}^{-\Gamma q^2 |t|}\cos\left(qc_s|t|\right)+\frac{3\Gamma-D_v}{\gamma\,  c_s}\, q \, {\rm e}^{-\Gamma q^2 |t|}\sin\left(qc_s|t|\right) .
\end{align}
Finally, taking the Fourier transform (\ref{eq:dsf}), the expression~(\ref{eq:dsf_hydro}) is found for the dynamic structure factor in the hydrodynamic approximation.  A similar but simpler calculation gives the spectral function~(\ref{eq:tvsf_hydro}) associated with the transverse components of velocity \cite{BP76,BY80}.

\section{Enskog approximation for the transport coefficients}
\label{app:Enskog}

Enskog's theory is an extension of Boltzmann's kinetic theory from dilute to dense gases \cite{B75,RD77}.  The system is assumed to have a positive second virial coefficient $B_2>0$.  The deviations of the equation of state for pressure with respect to ideality are taken into account with the function
\begin{align}\label{eq:y_E}
y_{\rm E} = \frac{p(n,T)}{n k_{\rm B} T} -1 = \frac{f(n)}{n}-1 \, ,
\end{align}
which is here evaluated with the truncated virial expansion~(\ref{eq:Virial_eos}).

Using Enskog's modification of Boltzmann's kinetic equation, the transport coefficients of the hard-sphere system where $B_2=2\pi d^3/3$ are given by~\cite{RD77,HCB54,SH03}
\begin{align}
\eta_{\rm E} &= B_2 n \left( \frac{1}{y_{\rm E}} + \frac{4}{5} + 0.76125 \, y_{\rm E} \right) \eta_{\rm B} \, , \\
\zeta_{\rm E} &= B_2 n \left( 1.01859\,  y_{\rm E}\right) \eta_{\rm B} \, , \\
\kappa_{\rm E} &= B_2 n \left( \frac{1}{y_{\rm E}} + \frac{6}{5} +  0.757 \, y_{\rm E} \right) \kappa_{\rm B} \, , 
\end{align}
in terms of the low-density approximations of Boltzmann's kinetic theory for the shear viscosity and heat conductivity
\begin{align}
\eta_{\rm B} = (1+  0.014 ) \, \frac{5\, m}{16 \, d^2} \, \sqrt{\frac{k_{\rm B}T}{\pi \, m}}
\qquad\mbox{and}\qquad
\kappa_{\rm B} =(1+ 0.025 ) \frac{75 \, k_{\rm B}}{64 \, d^2} \, \sqrt{\frac{k_{\rm B}T}{\pi \, m}} \, .
\end{align}


\newpage

\begin{table}[h!]
  \begin{tabular}{c @{\hskip 1cm}   c @{\hskip .2cm}  c  @{\hskip .2cm}  c}
    \hline\hline
    $n_*$ & $p_*$ & $p_{*{\rm V}}$ & $p_{*{\rm P}}$\\
    \hline
    0.1 &0.12397$\pm$0.00001& 0.12397 & 0.12397\\
    0.2 &0.31079$\pm$0.00002 &  0.31070&0.31072\\ 
    0.3 &0.59069$\pm$0.00008&0.59016 & 0.59047\\
    0.4 &1.0092$\pm$0.0002&1.0061 & 1.0086\\
    0.5 &1.6359$\pm$0.0003&1.6217&1.6346\\
    0.6 &2.5796$\pm$0.0006& 2.5254 & 2.5768\\
    0.7 &4.0132$\pm$0.0009&3.8394 & 4.0085\\
    0.8 &6.224$\pm$0.001&5.727 & 6.216\\
    0.9 &9.702$\pm$0.002&8.405&9.692\\
    \hline\hline
  \end{tabular}
  \caption{Pressure $p$ versus the density $n_*$, as computed with equation~\eqref{eq:pressureHS} using molecular dynamics simulation for a fluid of $N=500$ hard spheres and compared to the values predicted by the truncated virial expansion~\eqref{eq:Virial_eos} and the Pad\'e approximant~\eqref{eq:Padé_eos}. The error on the computed values is estimated as the difference between the values obtained by averaging over the whole time interval and its second half.}\label{Tab:P}
\end{table}


\begin{table}[h!]
  \begin{tabular}{c @{\hskip 1cm}c@{\hskip .2cm}c@{\hskip .2cm}c @{\hskip 1cm}c@{\hskip .2cm}c@{\hskip .2cm}c @{\hskip 1cm}c@{\hskip .2cm}c@{\hskip .2cm}c @{\hskip 1cm}c@{\hskip .2cm}c@{\hskip .2cm}c }
    \hline\hline
    $n_*$ & $c_{s*}$ & $c_{s*{\rm V}}$ & $c_{s*{\rm P}}$ & $c_{p*}$ & $c_{p*{\rm V}}$ & $c_{p*{\rm P}}$ & $\gamma$ & $\gamma_{\rm V}$ & $\gamma_{\rm P}$  & $\chi_{T*}$& $\chi_{T*{\rm V}}$ & $\chi_{T*{\rm P}}$   \\ 
    \hline
    0.1 & 1.592 & 1.593 & 1.593 & 2.516 & 2.516 & 2.516 & 1.677 & 1.677 & 1.677 & 6.609 & 6.610 & 6.610 \\
    0.2 & 1.970 & 1.970 & 1.970 & 2.562 & 2.563 & 2.563 & 1.708 & 1.708 & 1.708 & 2.200 & 2.202 & 2.201   \\
    0.3 & 2.446 & 2.443 & 2.445 & 2.640 & 2.641 & 2.640 & 1.760 & 1.761 & 1.760 & 0.980 & 0.983 & 0.981  \\
    0.4 & 3.054 & 3.042 & 3.053 & 2.752 & 2.757 & 2.752 & 1.835 & 1.838 & 1.835 & 0.491 & 0.497 & 0.492 \\
    0.5 & 3.842 & 3.800 & 3.840 & 2.904 & 2.917 & 2.903 & 1.936 &1.945 &1.935 &  0.262 & 0.269 & 0.263\\
    0.6 & 4.883 & 4.761 & 4.879 & 3.104 & 3.132 & 3.103 &  2.069 & 2.088 & 2.069 & 0.145 & 0.154 & 0.145\\
    0.7 & 6.287 & 5.979 & 6.281 & 3.366 & 3.417 & 3.364 & 2.244 & 2.278 & 2.243 & 0.0811 & 0.0910 & 0.0812 \\
    0.8 & 8.230 & 7.523 & 8.222 & 3.710 & 3.786 & 3.707 & 2.473 & 2.524 & 2.471 & 0.0456 & 0.0558 & 0.0457 \\
    0.9 & 11.003 & 9.473 & 10.994 & 4.166 & 4.260 & 4.161 &   2.777 & 2.840 & 2.774 & 0.0255 & 0.0352 & 0.0255
     \\      \hline\hline
  \end{tabular}
  \caption{Thermodynamic properties versus the density $n_*$:  The adiabatic speed of sound $c_s$ is given by equation~\eqref{eq:c_s}, the specific heat capacity $c_p$ by equation~\eqref{eq:c_P}, the specific heat ratio $\gamma$ by equation~\eqref{eq:gamma}, and the isothermal compressibility $\chi_T$ by equation~\eqref{eq:chi_T}, as computed using molecular dynamics simulation for a fluid of $N=500$ hard spheres and compared to the values predicted by the truncated virial expansion~\eqref{eq:Virial_eos} and  the Pad\'e approximant~\eqref{eq:Padé_eos}. }\label{Tab:TP}
\end{table}


\begin{table}[h!]
  \begin{tabular}{c @{\hskip 1cm} c @{\hskip .2cm}c @{\hskip 1cm} c c}
    \hline\hline
    $n_*$ & $ \lim_{q\rightarrow 0}[S(q)]_*$ (Lin. Reg.) & $\lim_{q\rightarrow 0}[S(q)]_*$ (PY) &$\chi_{T*}$&  $\chi_{T*{\rm PY}}$      \\ \hline
    0.1 & 0.665$\pm$0.001 & 0.661 & 6.646 & 6.608 \\
    0.2 & 0.443$\pm$0.002 & 0.439 & 2.216 & 2.196 \\
    0.3 & 0.295$\pm$0.002 & 0.292 & 0.983 & 0.974 \\
    0.4 & 0.196$\pm$0.002 & 0.194 & 0.489 & 0.485 \\
    0.5 & 0.129$\pm$0.002 & 0.128 & 0.258 & 0.256 \\
    0.6 & 0.085$\pm$0.002 & 0.083 & 0.141 & 0.139 \\
    0.7 & 0.055$\pm$0.001 & 0.054 & 0.078 & 0.077 \\
    0.8 & 0.035$\pm$0.001 & 0.034 & 0.044 & 0.042 \\
    0.9 & 0.022$\pm$0.001 & 0.021 & 0.024 & 0.023 \\
          \hline \hline
  \end{tabular}
  \caption{Limit $q\rightarrow 0$ of the static structure factor versus the density $n_*$.  The limit $q\to 0$ is obtained using linear regression with standard error on the fitted parameter. PY corresponds to the Percus-Yervick approximation.  The corresponding isothermal compressibilities are given by equation~\eqref{eq:isocompSq}. }\label{Tab:SSF}
\end{table}


\begin{table}[h!]
  \begin{tabular}{ c @{\hskip 1cm} c @{\hskip 0.3cm} c @{\hskip 1cm}c @{\hskip 0.3cm} c  @{\hskip 1cm}c @{\hskip 0.3cm} c @{\hskip 1cm} c @{\hskip 0.3cm}  c  }
    \hline\hline
     $n_*$ & $\eta_*$ & $\eta_{*{\rm E}}$  & $\zeta_*$ & $\zeta_{*{\rm E}}$ & $\zeta_* + \frac{4}{3}\eta_*$ & $\zeta_{*{\rm E}} + \frac{4}{3}\eta_{*{\rm E}}$ & $\kappa_*$ & $\kappa_{*{\rm E}}$\\ \hline
     0.1 & 0.193$\pm$0.001 & 0.193 & 0.009$\pm$0.001 & 0.009 & 0.265$\pm$0.001 & 0.266 & 0.771$\pm$0.005 & 0.788\\
     0.2 & 0.234$\pm$0.004 & 0.227 & 0.041$\pm$0.001 & 0.042 & 0.338$\pm$0.005 & 0.345 & 0.960$\pm$0.006 & 0.972\\
     0.3 & 0.300$\pm$0.002 & 0.289 & 0.114$\pm$0.004 & 0.111 & 0.51$\pm$0.02 & 0.496 & 1.26$\pm$0.04 & 1.26\\
     0.4 & 0.393$\pm$0.008 & 0.391 & 0.230$\pm$0.003 & 0.231 & 0.741$\pm$0.001 & 0.753 & 1.723$\pm$0.005 & 1.707\\
     0.5 & 0.58$\pm$0.02 & 0.55 & 0.431$\pm$0.001 & 0.428 & 1.178$\pm$0.003 & 1.165 & 2.38$\pm$0.08 & 2.37\\
     0.6 & 0.79$\pm$0.02 & 0.80 & 0.73$\pm$0.02 & 0.73 & 1.79$\pm$0.05 & 1.80 & 3.42$\pm$0.05 & 3.36\\
     0.7 & 1.26$\pm$0.04 & 1.16 & 1.26$\pm$0.03 & 1.20 & 2.90$\pm$0.04 & 2.75 & 4.96$\pm$0.1 & 4.79\\
     0.8 & 2.18$\pm$0.05 & 1.69 & 2.22$\pm$0.01 & 1.88 & 5.1$\pm$0.1 & 4.1 & 7.52$\pm$0.04 & 6.84\\
     0.9 & 4.52$\pm$0.09 & 2.45 & 5.1$\pm$0.2 & 2.9 & 11.0$\pm$0.2 & 6.1 & 11.45$\pm$0.07 & 9.75\\
  \hline\hline
  \end{tabular}
    \caption{Transport coefficients versus the density $n_*$: The shear viscosity $\eta$ is given by equation~\eqref{eq:etaH}, the bulk viscosity $\zeta$ by equation~\eqref{eq:zetaH}, the longitudinal viscosity $\zeta+\frac{4}{3}\eta$ by equation~\eqref{eq:longviscH}, and the heat conductivity $\kappa$ by equation~\eqref{eq:kappaH} using molecular dynamics simulation for a fluid of $N=500$ hard spheres. The errors are estimated as the difference with the same coefficient obtained from a regression over the second half of the trajectory.  The Enskog approximations to the transport coefficients are calculated as explained in appendix~\ref{app:Enskog}.}\label{Tab:TC}
\end{table}

\begin{table}[h!]
  \begin{tabular}{c @{\hskip 1cm} c @{\hskip 1cm} c @{\hskip 1cm} c @{\hskip 1cm} c}
    \hline\hline
    $q_*$  & $ [c_s(q)]_*$   &      $[\Gamma(q)]_*$       &     $[D_{T}(q)]_*$ &  $[\eta(q)]_*$	\\
    \hline     
    0.82 & 1.86$\pm$0.02 & 1.28$\pm$0.01 & 1.53$\pm$0.03& 0.197$\pm$0.009 \\
    0.78 & 1.84$\pm$0.01 & 1.20$\pm$0.01 & 1.59$\pm$0.02 & 0.183$\pm$0.009 \\
    0.73 & 1.82$\pm$0.01 & 1.34$\pm$0.01 & 1.56$\pm$0.01 & 0.199$\pm$0.007 \\
    0.69 & 1.82$\pm$0.02 & 1.44$\pm$0.01 & 1.81$\pm$0.01 & 0.185$\pm$0.004 \\
    0.64 & 1.84$\pm$0.03 & 1.52$\pm$0.02 & 1.77$\pm$0.05 & 0.190$\pm$0.003 \\
    0.58 & 1.78$\pm$0.02 & 1.48$\pm$0.02 & 1.81$\pm$0.02 & 0.202$\pm$0.003 \\
    0.52 & 1.84$\pm$0.02 & 1.62$\pm$0.02 & 1.96$\pm$0.01 & 0.203$\pm$0.002 \\
    0.45 & 1.78$\pm$0.02 & 1.75$\pm$0.02 & 2.19$\pm$0.03 & 0.215$\pm$0.002 \\
    0.37 & 1.79$\pm$0.01 & 1.73$\pm$0.02 & 2.26$\pm$0.03 & 0.201$\pm$0.003 \\
    0.26 & 1.80$\pm$0.01 & 1.78$\pm$0.02 & 2.37$\pm$0.01 & 0.207$\pm$0.001 \vspace{0.1cm}\\
\vspace{0.1cm}
    0 (Lin. Reg.) & 1.78$\pm$0.01 & 1.87$\pm$0.04 & 2.41 $\pm $0.05 & 0.21$\pm$0.01\\
    0 (Helfand) & 1.75 & 1.88 & 2.42 & 0.21\\
    \hline\hline
  \end{tabular}
  \caption{Dependence on the wave number $q$ for the speed of sound and the transport coefficients obtained from the poles of the correlation functions $S(q,\omega)$ and $J_{\rm t}(q,\omega)$ given by the numerical Fourier transform of the intermediate scattering function~\eqref{eq:isf} and the transverse component of the correlation function~\eqref{eq:vcf} at density $n_*=0.144$ for a fluid of $N=2048$ hard spheres. The reported error is on the estimation of the location of the pole. The penultimate row (Lin. Reg.) corresponds to the limit $q\rightarrow 0$ and is obtained using a linear least square  regression with equal weights on the data points for the different values of $q$. The reported error is the standard error on the fitted parameter. The last row (Helfand) is the corresponding value obtained from the Helfand moments.}\label{Tab:CC0144}
\end{table}

\begin{table}[h!]
  \begin{tabular}{c @{\hskip 1cm} c @{\hskip 1cm} c @{\hskip 1cm} c @{\hskip 1cm} c}
    \hline\hline
    $q_*$  & $[c_s(q)]_*$   &      $[\Gamma(q)]_*$       &     $[D_T(q)]_*$ &  $[\eta(q)]_*$	\\
    \hline   
    1.95& 2.67 $\pm$0.03 & 1.16$\pm$0.02 & 1.00$\pm$0.02 & 0.354$\pm$0.01 \\
    1.84& 2.32 $\pm$0.05 & 1.24$\pm$0.04& 1.05$\pm$0.04 & 0.365$\pm$0.003 \\
    1.74& 2.58 $\pm$0.01 & 1.27$\pm$0.01 & 1.12$\pm$0.03 & 0.403$\pm$0.008 \\
    1.63& 2.65 $\pm$0.03& 1.35$\pm$0.01 & 1.15$\pm$0.02 & 0.400$\pm$0.007 \\
    1.51 & 2.97 $\pm$0.04 & 1.35$\pm$0.01 & 1.34$\pm$0.06 & 0.407$\pm$0.02 \\
    1.38 & 3.18 $\pm$0.03 & 1.55$\pm$0.03 & 1.39$\pm$0.02 & 0.415$\pm$0.003 \\
    1.23 & 3.21 $\pm$0.02 & 1.51$\pm$0.01 & 1.43$\pm$0.03 & 0.437$\pm$0.003 \\
    1.07 & 3.07 $\pm$0.01 & 1.60$\pm$0.03& 1.36$\pm$0.08& 0.442$\pm$0.005 \\
    0.87 & 3.33 $\pm$0.01 & 1.64$\pm$0.01 & 1.42$\pm$0.02 & 0.467$\pm$0.007 \\
    0.62 & 3.56 $\pm$0.01 & 1.75$\pm$0.02 & 1.60$\pm$0.03 & 0.485$\pm$0.002 \vspace{0.1cm}\\
    \vspace{0.1cm}
    0 (Lin. Reg.) & 3.60$\pm$0.1 & 1.79$\pm$0.03 & 1.62 $\pm $0.04 & 0.491$\pm$0.006\\
    0 (Helfand) & 3.59 & 1.83 & 1.61 & 0.498\\
    \hline\hline
      \end{tabular}
  \caption{Dependence on the wave number $q$ for the speed of sound and the transport coefficients obtained from the poles of the correlation functions $S(q,\omega)$ and $J_{\rm t}(q,\omega)$ given by the numerical Fourier transform of the intermediate scattering function~\eqref{eq:isf} and the transverse component of the correlation function~\eqref{eq:vcf} at density $n_*=0.471$ for a fluid of $N=500$ hard spheres. The reported error is on the estimation of the location of the pole.  The penultimate row (Lin. Reg.) corresponds to the limit $q\rightarrow 0$ and is obtained using a linear least square  regression with equal weights on the data points for the different values of $q$. The reported error is the standard error on the fitted parameter. The last row (Helfand) is the corresponding value obtained from the Helfand moments.}\label{Tab:CC0471}
\end{table}

\begin{table}[h!]
  \begin{tabular}{c @{\hskip 1cm} c @{\hskip 1cm} c @{\hskip 1cm} c @{\hskip 1cm} c}
    \hline\hline
    $q_*$  & $[c_s(q)]_*$   &      $[\Gamma(q)]_*$       &     $[D_T(q)]_*$ &  $[\eta(q)]_*$	\\
    \hline   
    2.40& 5.0 $\pm$0.3 & 3.67$\pm$0.07 & 0.51$\pm$0.01 & 2.85$\pm$0.04 \\
    2.28& 6.30 $\pm$0.01 & 3.86$\pm$0.02& 0.75$\pm$0.02 & 2.54$\pm$0.04 \\
    2.15& 6.1 $\pm$0.1 & 3.7$\pm$0.2 & 0.87$\pm$0.09 & 2.88$\pm$0.02 \\
    2.01& 5.6 $\pm$0.1& 3.29$\pm$0.04 & 0.87$\pm$0.03 &2.8$\pm$0.1 \\
    1.86 & 6.3 $\pm$0.3 & 3.7$\pm$0.1 & 0.94$\pm$0.09 & 3.0$\pm$0.1 \\
    1.70 & 7.39 $\pm$0.08 & 4.37$\pm$0.03 & 1.09$\pm$0.04 & 3.24$\pm$0.01 \\
   1.52 & 7.85 $\pm$0.07 & 4.57$\pm$0.05 & 1.31$\pm$0.05 & 3.3$\pm$0.1 \\
    1.32 & 8.44 $\pm$0.05 & 5.32$\pm$0.05& 1.8$\pm$0.1& 3.29$\pm$0.09 \\
    1.07 & 9.15 $\pm$0.04 & 6.3$\pm$0.1 & 2.08$\pm$0.07 & 3.509$\pm$0.004 \\
    0.76 & 10.15 $\pm$0.02 & 7.2$\pm$0.1 & 2.53$\pm$0.05 & 3.768$\pm$0.005 \vspace{0.1cm}\\
    \vspace{0.1cm}
    0 (Lin. Reg.) & 10.7$\pm$0.2 & 7.6$\pm$0.2 & 2.8$\pm $0.1 & 3.80$\pm$0.08\\
    0 (Helfand) & 10.5 & 7.8 & 3.0 & 3.87\\
    \hline\hline 
      \end{tabular}
  \caption{Dependence on the wave number $q$ for the speed of sound and the transport coefficients obtained from the poles of the correlation functions $S(q,\omega)$ and $J_{\rm t}(q,\omega)$ given by the numerical Fourier transform of the intermediate scattering function~\eqref{eq:isf} and the transverse component of the correlation function~\eqref{eq:vcf} at density $n_*=0.884$ for a fluid of $N=500$ hard spheres. The reported error is on the estimation of the location of the pole. The penultimate row (Lin. Reg.) corresponds to the limit $q\rightarrow 0$ and is obtained with a linear least square regression with equal weights on the data points for the different values of $q$. The reported error is the standard error on the fitted parameter. The last row (Helfand) is the corresponding value obtained from the Helfand moments.}\label{Tab:CC0884}
\end{table}


\begin{figure}[h!]\centering
{\includegraphics[width=0.95\textwidth]{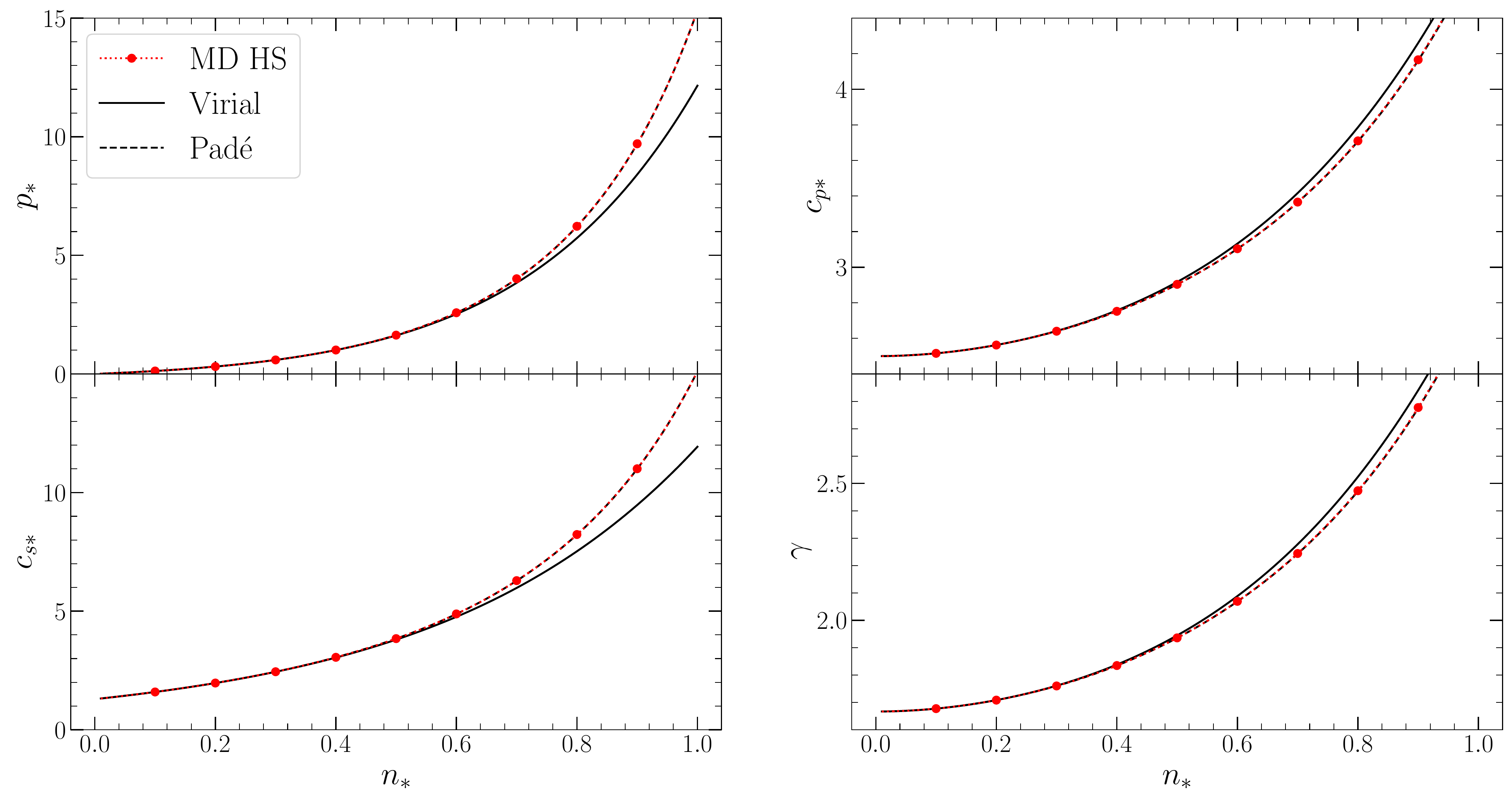}}
\caption[] {Thermodynamic properties versus the density $n_*$: The pressure $p$ is given by equation~\eqref{eq:pressureHS}, the speed of sound $c_s$ by equation~\eqref{eq:c_s}, the specific heat capacity at constant pressure $c_p$ by equation~\eqref{eq:c_P}, and the specific heat ratio $\gamma$ by equation~\eqref{eq:gamma}, as computed using the molecular dynamic simulation for a fluid of $N=500$ hard spheres, and compared to the values predicted by the truncated virial expansion~\eqref{eq:Virial_eos} and  the Pad\'e approximant~\eqref{eq:Padé_eos}.}\label{Fig:TP}
\end{figure}


\begin{figure}[h!]\centering
{\includegraphics[width=0.55\textwidth]{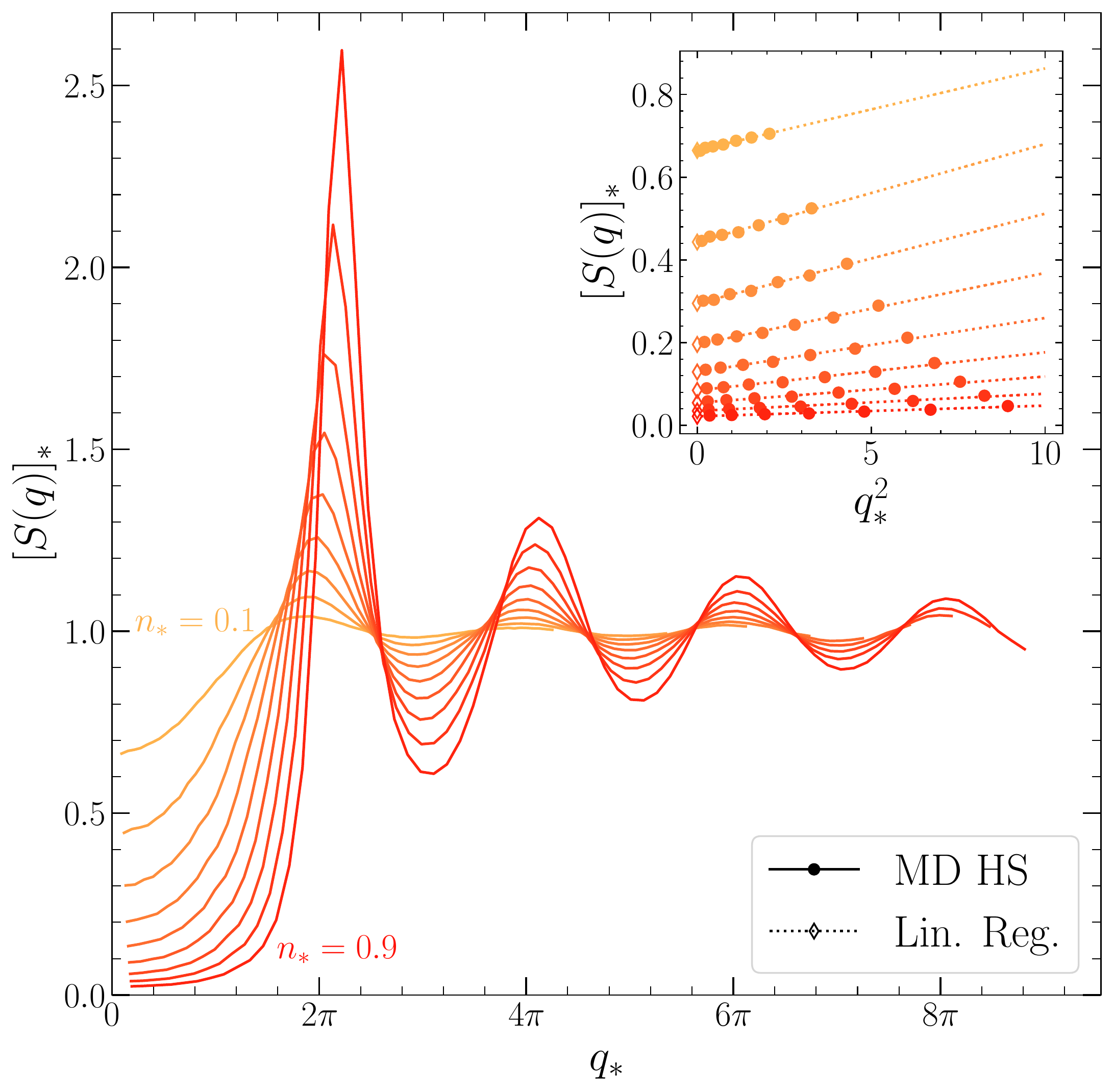}}
\caption[] {Static structure factor~\eqref{eq:ssf} for a fluid of $500$ hard spheres at the densities $n_* = 0.1,...,0.9$.
Inset: Small-$q$ limit of the static structure factor (filled circles) versus $q^2$, linear regression approximation (dotted lines), and extrapolation to $q=0$ (open diamonds).
}\label{Fig:SSF}
\end{figure}


\begin{figure}[h!]\centering
{\includegraphics[width=0.9\textwidth]{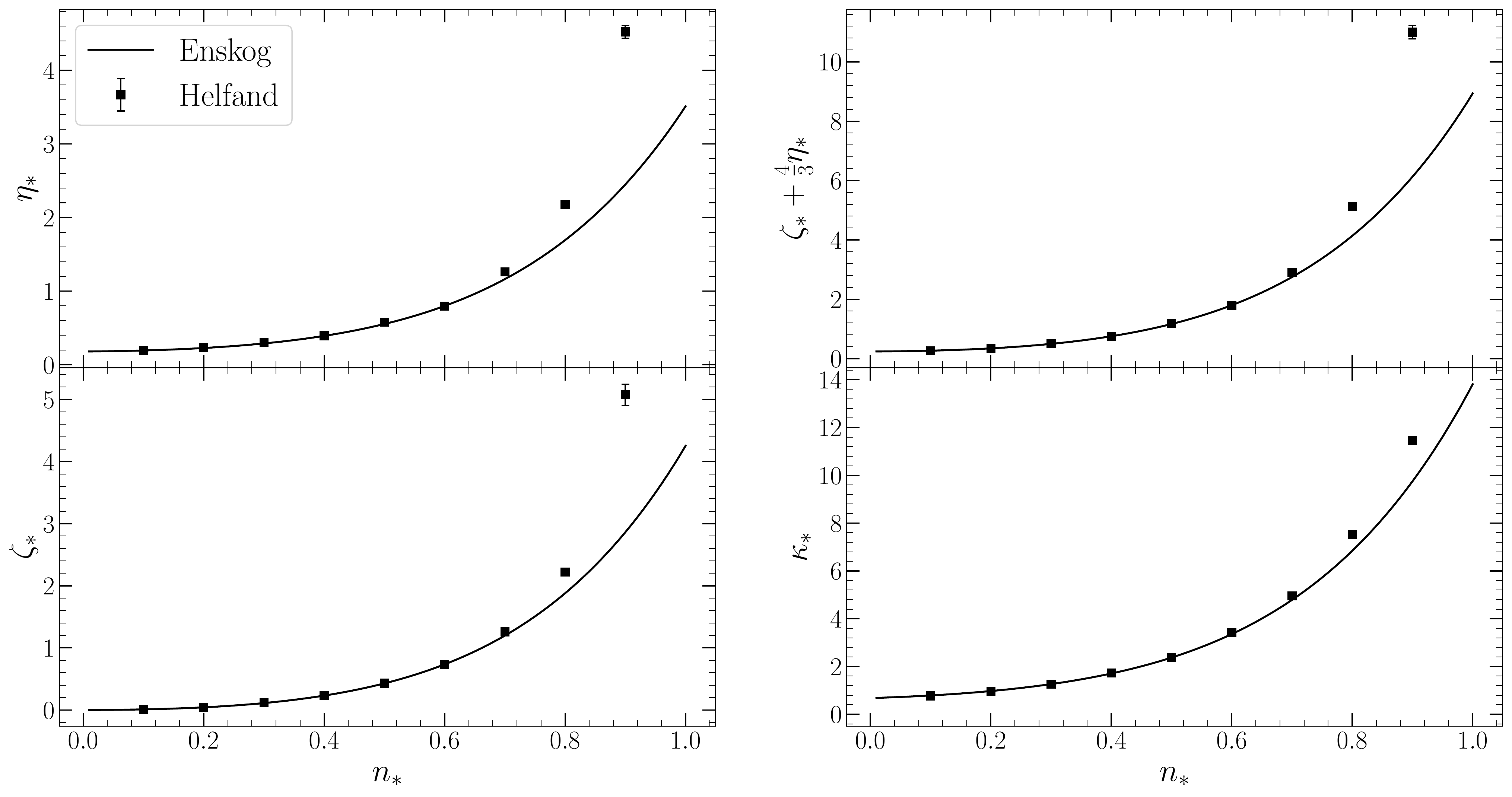}}
\caption[] {Transport coefficients versus the density $n_*$: The shear viscosity $\eta$ is given by equation~\eqref{eq:etaH}, the bulk viscosity $\zeta$ by equation~\eqref{eq:zetaH}, the longitudinal viscosity $\zeta+\frac{4}{3}\eta$ by equation~\eqref{eq:longviscH}, and the heat conductivity $\kappa$ by equation~\eqref{eq:kappaH}, as computed using molecular dynamics simulation for a fluid of $N=500$ hard spheres, and compared to the Enskog predictions of appendix~\ref{app:Enskog}. When not appearing, the error bars are within the size of the symbols.
}\label{Fig:TC}
\end{figure}


\begin{figure}[h!]\centering
{\includegraphics[width=1.0\textwidth]{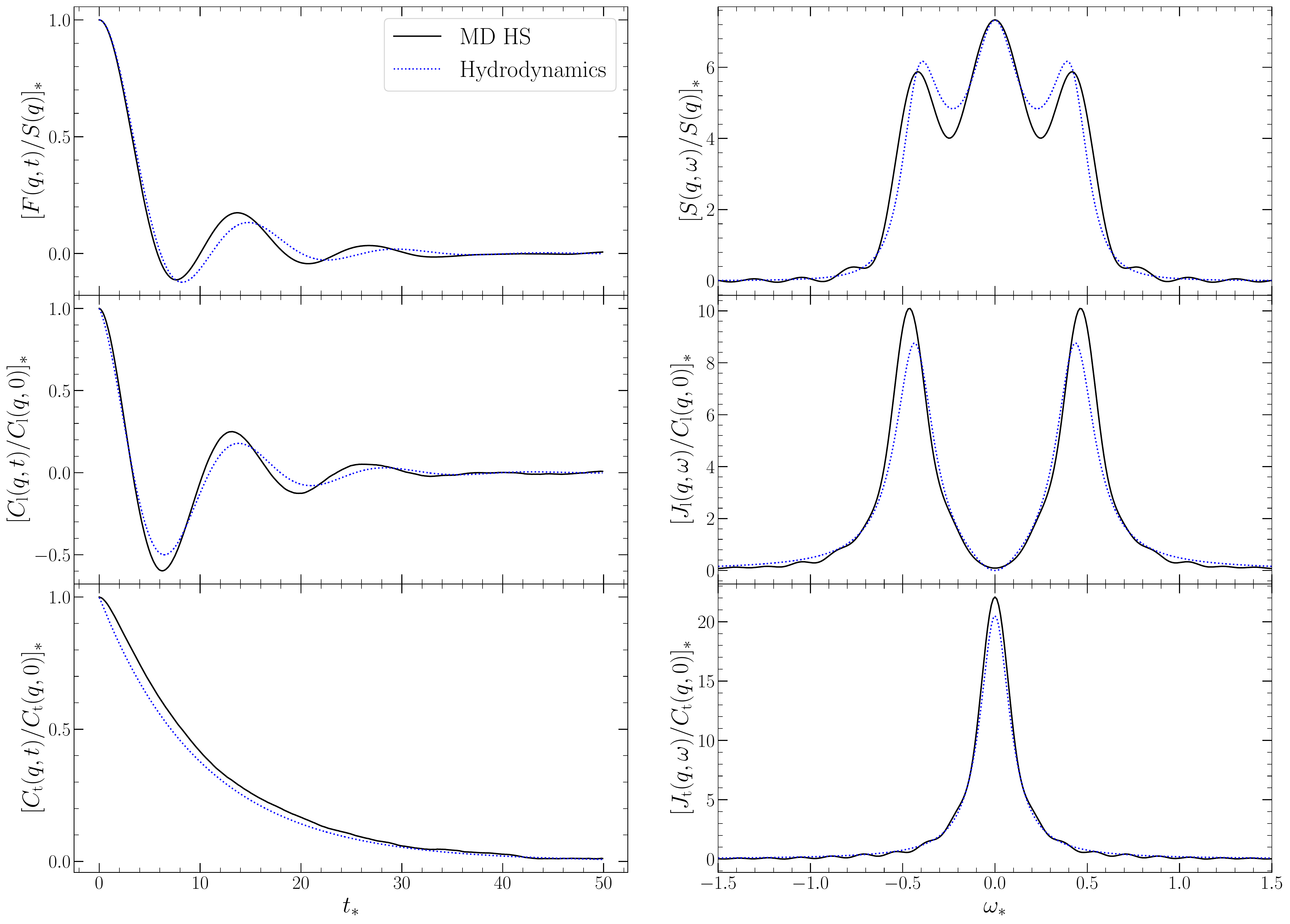}}
\caption[] {Normalized correlation and spectral functions at density $n_*=0.144$ and $q_*=0.259$ for a fluid of $N=2048$ hard spheres. Left panel: Correlation functions versus time: From top to bottom: Intermediate scattering function $F(q,t)$ given by equation~\eqref{eq:isf}, longitudinal component $C_{\rm l}(q,t)$, and transverse component $C_{\rm t}(q,t)$ of the time-dependent correlation functions~\eqref{eq:vcf}, normalized by their values at $t=0$. Right panel: Spectral functions versus frequency:  From top to bottom: Dynamic structure factor $S(q,\omega)$, longitudinal component $J_{\rm l}(q,\omega)$, and transverse component $J_{\rm t}(q,\omega)$ of the spectral functions~\eqref{eq:vsf} obtained from a numerical Fourier transform of the corresponding time-dependent correlation functions.  The dotted lines correspond to the hydrodynamic approximations given by equations~\eqref{eq:dsf_hydro}, ~\eqref{eq:lca_hydro}, and ~\eqref{eq:tvsf_hydro} and their temporal Fourier transforms.}\label{Fig:CC0144}
\end{figure}


\begin{figure}[h!]\centering
{\includegraphics[width=1.0\textwidth]{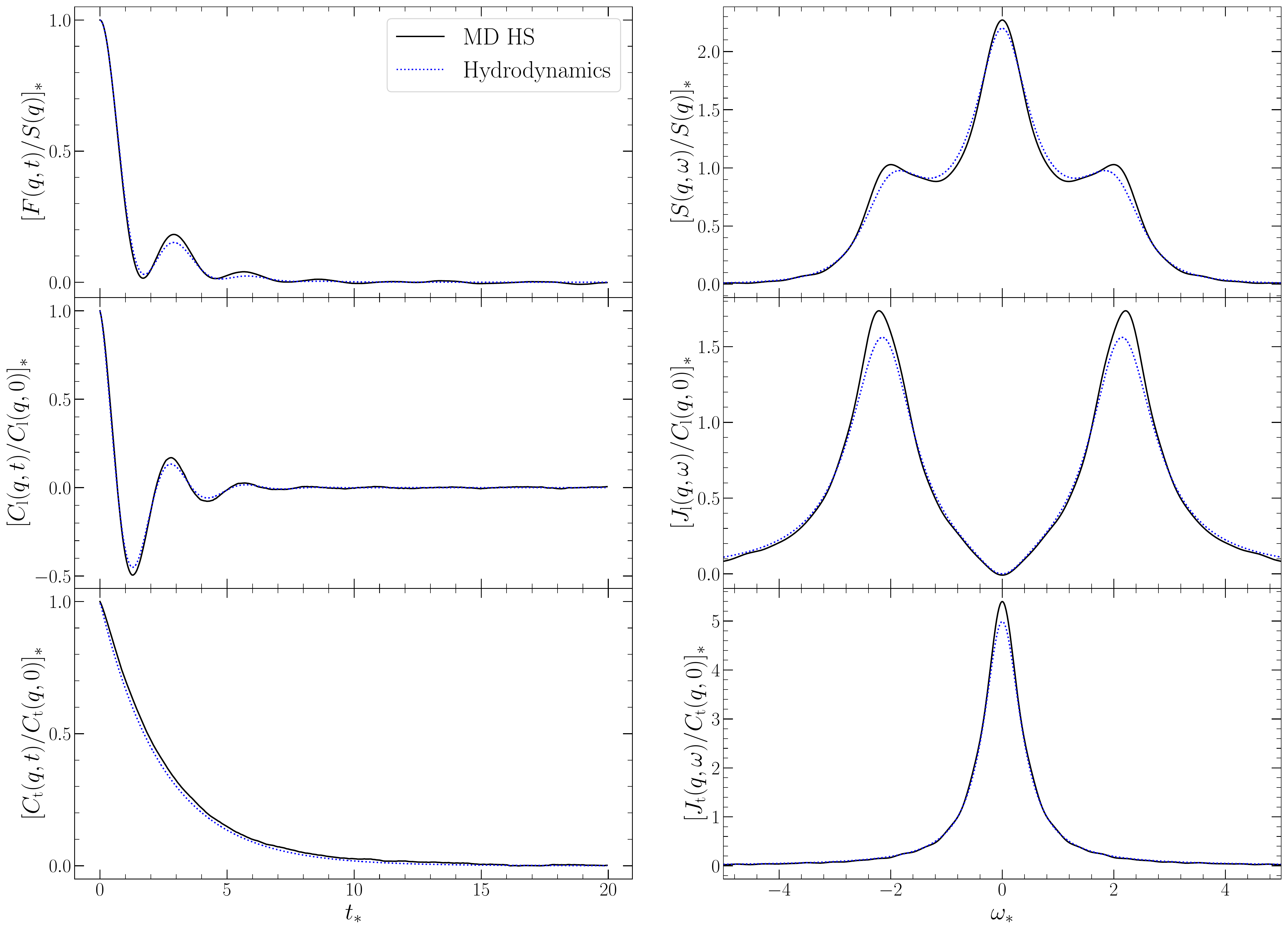}}
\caption[] {Normalized correlation and spectral functions at density $n_*=0.471$ and $q_*=0.616$ for a fluid of $N=500$ hard spheres. Left panel: Correlation functions versus time: From top to bottom: Intermediate scattering function $F(q,t)$ given by equation~\eqref{eq:isf}, longitudinal component $C_{\rm l}(q,t)$, and transverse component $C_{\rm t}(q,t)$ of the time-dependent correlation functions~\eqref{eq:vcf}, normalized by their values at $t=0$. Right panel: Spectral functions versus frequency:  From top to bottom: Dynamic structure factor $S(q,\omega)$, longitudinal component $J_{\rm l}(q,\omega)$, and transverse component $J_{\rm t}(q,\omega)$ of the spectral functions~\eqref{eq:vsf} obtained from a numerical Fourier transform of the corresponding time-dependent correlation functions.  The dotted lines correspond to the hydrodynamic approximations given by equations~\eqref{eq:dsf_hydro}, ~\eqref{eq:lca_hydro}, and ~\eqref{eq:tvsf_hydro} and their temporal Fourier transforms.}\label{Fig:CC0471}
\end{figure}


\begin{figure}[h!]\centering
{\includegraphics[width=1.0\textwidth]{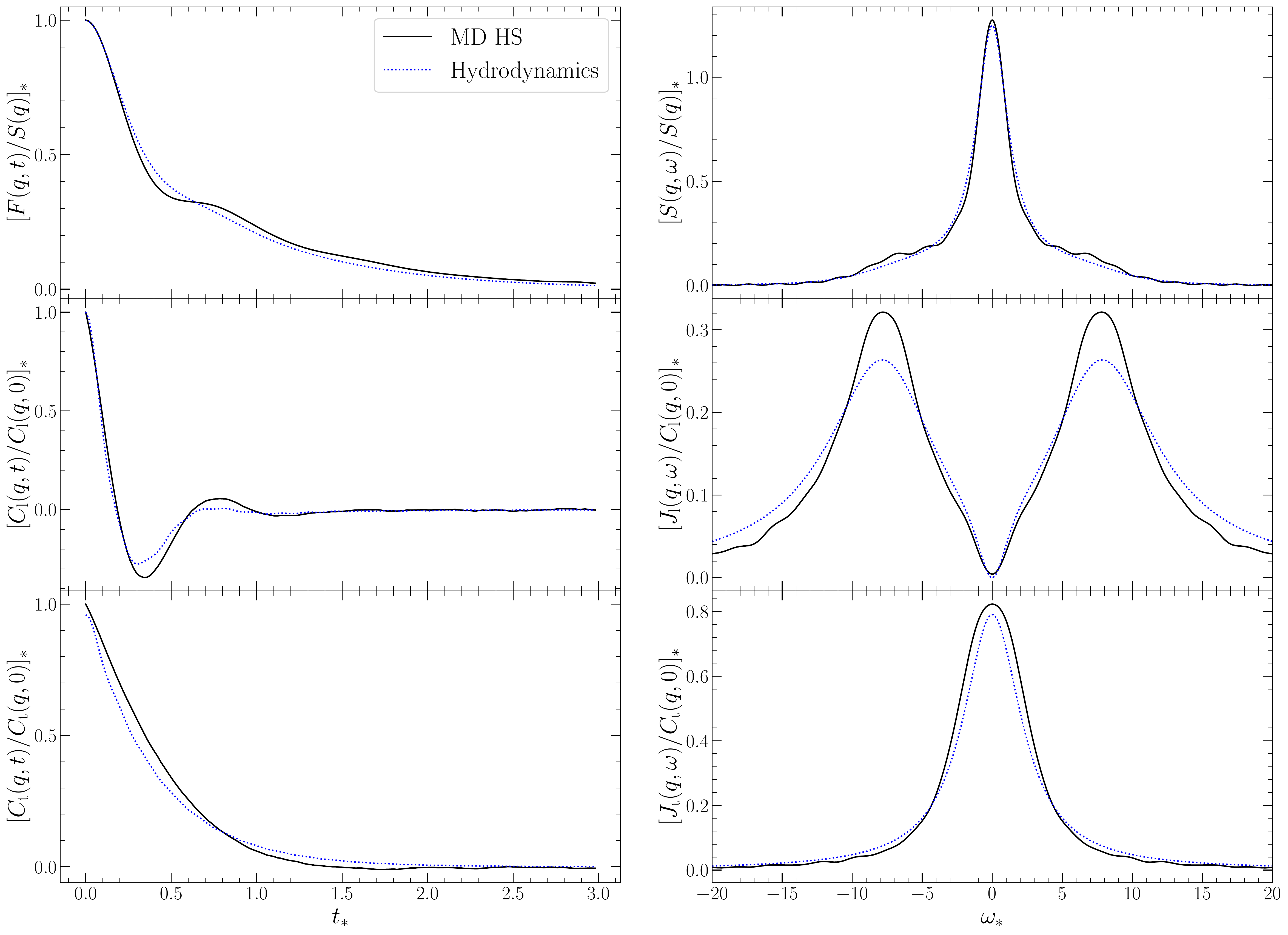}}
\caption[] {Normalized correlation and spectral functions at density $n_*=0.884$ and $q_*=0.760$ for a fluid of $N=500$ hard spheres. Left panel: Correlation functions versus time: From top to bottom: Intermediate scattering function $F(q,t)$ given by equation~\eqref{eq:isf}, longitudinal component $C_{\rm l}(q,t)$, and transverse component $C_{\rm t}(q,t)$ of the time-dependent correlation functions~\eqref{eq:vcf}, normalized by their values at $t=0$. Right panel: Spectral functions versus frequency:  From top to bottom: Dynamic structure factor $S(q,\omega)$, longitudinal component $J_{\rm l}(q,\omega)$, and transverse component $J_{\rm t}(q,\omega)$ of the spectral functions~\eqref{eq:vsf} obtained from a numerical Fourier transform of the corresponding time-dependent correlation functions.  The dotted lines correspond to the hydrodynamic approximations given by equations~\eqref{eq:dsf_hydro}, ~\eqref{eq:lca_hydro}, and ~\eqref{eq:tvsf_hydro} and their temporal Fourier transforms.}\label{Fig:CC0884}
\end{figure}


\begin{figure}[h!]\centering
{\includegraphics[width=0.52\textwidth]{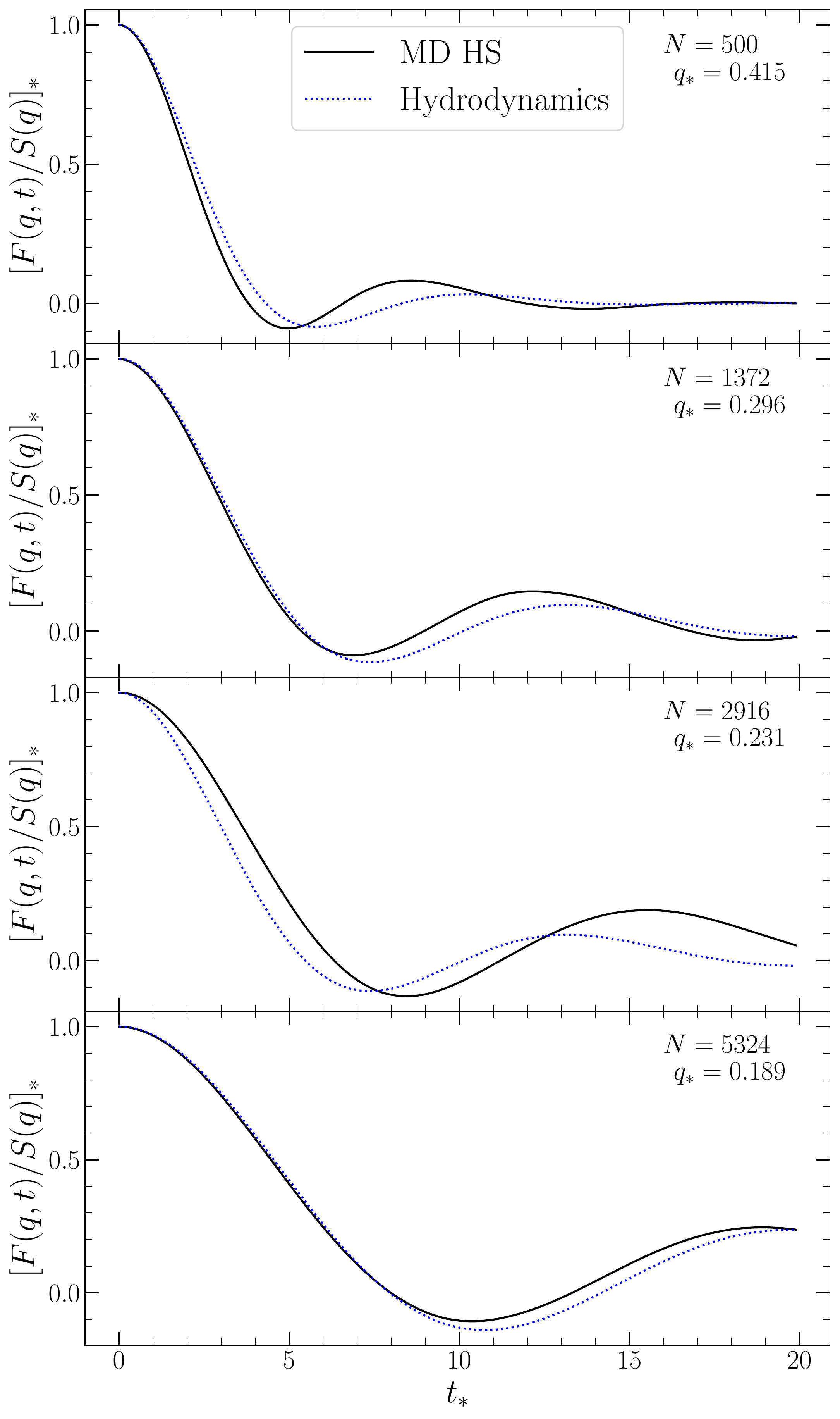}}
\caption[] {Intermediate scattering function $F(q,t)$ for the smallest value of $q$ and increasing particle number $N$, compared to the hydrodynamic expression given by the numerical Fourier transform of equation~\eqref{eq:dsf_hydro} in time.}\label{Fig:HL}
\end{figure}


\begin{figure}[h!]\centering
{\includegraphics[width=0.55\textwidth]{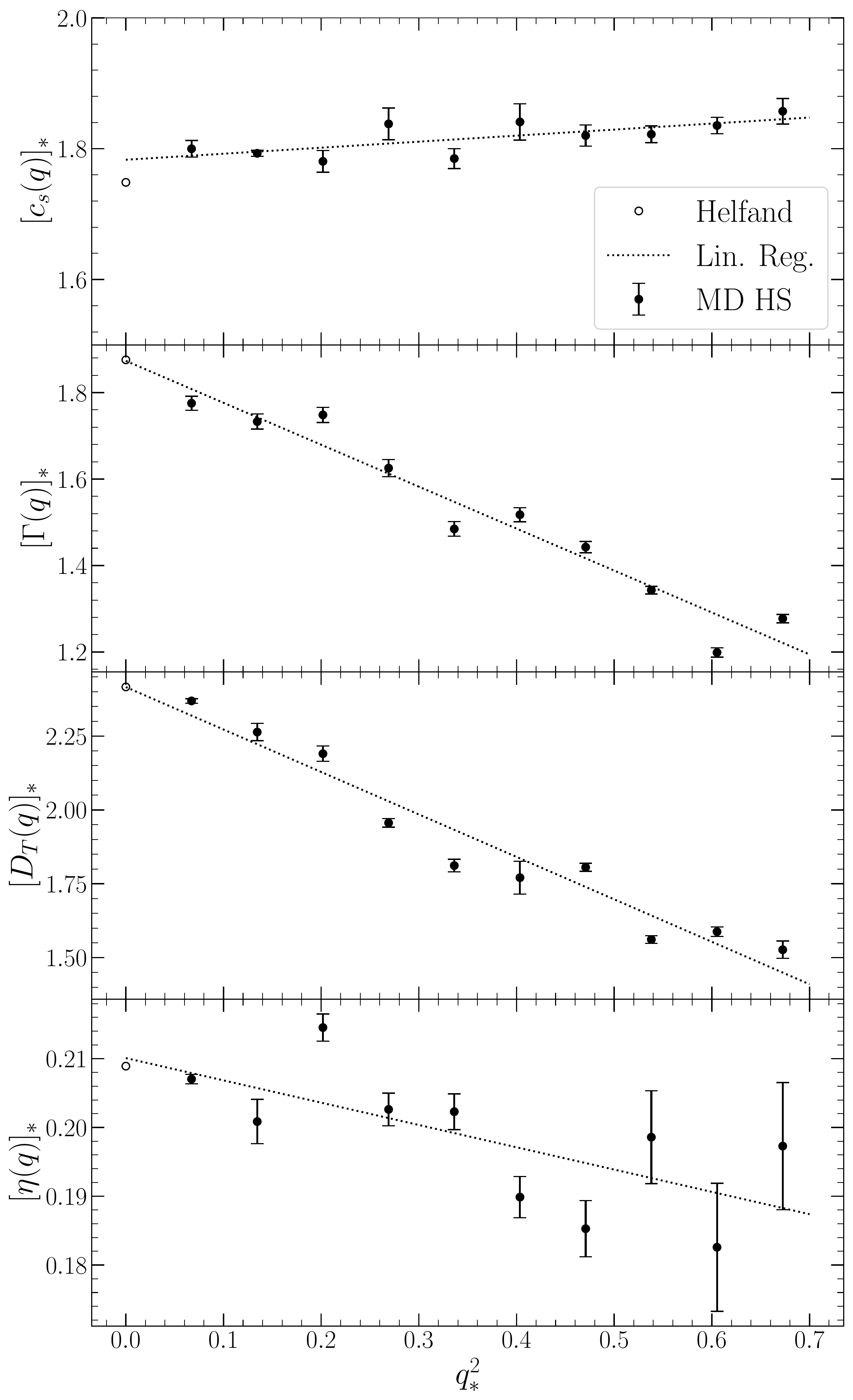}}
\caption[] {Extrapolation to $q=0$ for the speed of sound and the transport coefficients obtained from the poles of spectral functions $S(q,\omega)$ and $J_{\rm t}(q,\omega)$ given by the numerical Fourier transform of equations~\eqref{eq:isf} and~\eqref{eq:vcf}, at density $n_*=0.144$ for a fluid of $N=2048$ hard spheres. The dotted line corresponds to the linear least square regression. The value at $q=0$ shown with an open circle is obtained using the Helfand moments.}\label{Fig:LR0144}
\end{figure}


\begin{figure}[h!]\centering
{\includegraphics[width=0.55\textwidth]{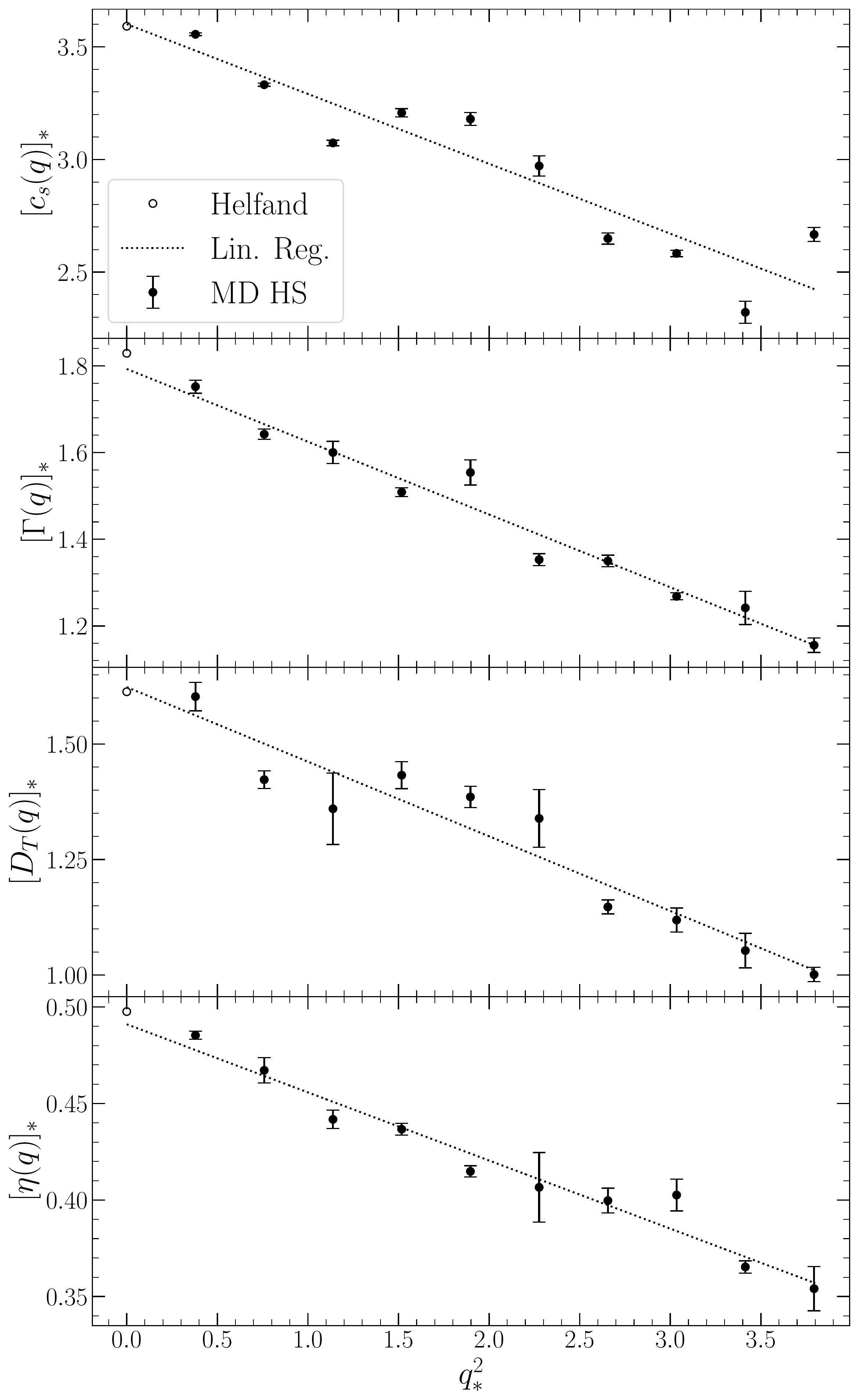}}
\caption[] {Extrapolation to $q=0$ for the speed of sound and the transport coefficients obtained from the poles of spectral functions $S(q,\omega)$ and $J_{\rm t}(q,\omega)$ given by the numerical Fourier transform of equations~\eqref{eq:isf} and~\eqref{eq:vcf}, at a density $n_*=0.471$ for a fluid of $N=500$ hard spheres. The dotted line corresponds to the linear least square  regression. The value at $q=0$ shown with an open circle is obtained using the Helfand moments.}\label{Fig:LR0471}
\end{figure}


\begin{figure}[h!]\centering
{\includegraphics[width=0.55\textwidth]{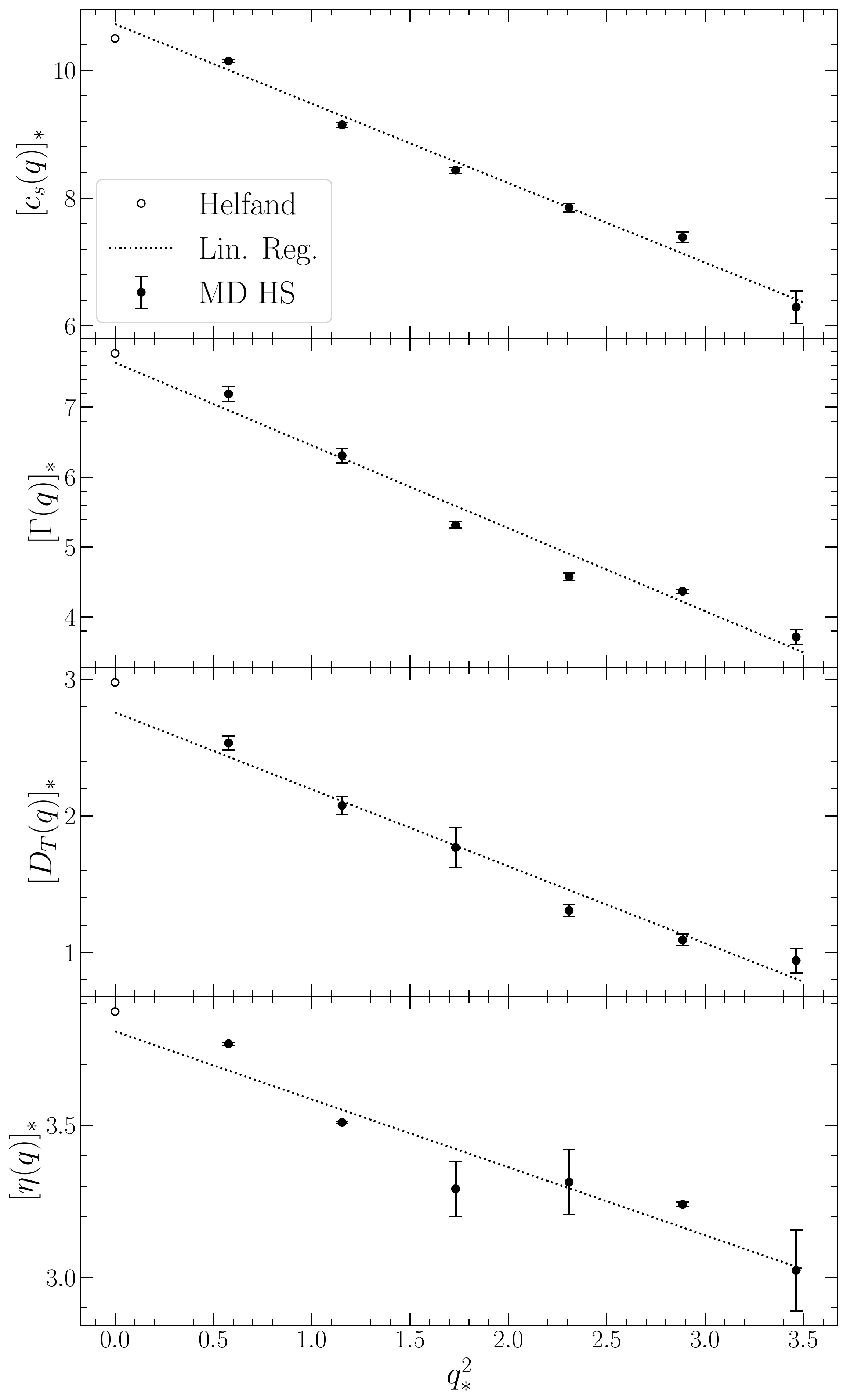}}
\caption[] {Extrapolation to $q=0$ for the speed of sound and the transport coefficients obtained from the poles of spectral functions $S(q,\omega)$ and $J_{\rm t}(q,\omega)$ given by the numerical Fourier transform of equations~\eqref{eq:isf} and~\eqref{eq:vcf}, at a density $n_*=0.884$ for a fluid of $N=500$ hard spheres. The dotted line corresponds to the linear least square  regression. The value at $q=0$ shown with an open circle is obtained using the Helfand moments.}\label{Fig:LR0884}
\end{figure}


\begin{thebibliography}{99}

\bibitem{GM84} S.~R.~de Groot and P.~Mazur, {\it Nonequilibrium Thermodynamics} (Dover, New York, 1984).

\bibitem{M58} H. Mori, {\it Statistical-Mechanical Theory of Transport in Fluids}, Phys. Rev. {\bf 112}, 1829-1842 (1958).

\bibitem{McL63} J. A. McLennan, {\it The Formal Statistical Theory of Transport Processes}, Adv. Chem. Phys. {\bf 5}, 261-317 (1963).

\bibitem{R66} B. Robertson, {\it Equations of Motion in Nonequilibrium Statistical Mechanics}, Phys. Rev. {\bf 144}, 151-161 (1966).

\bibitem{P68} R. A. Piccirelli, {\it Theory of the dynamics of simple fluids for large spatial gradients and long memory}, Phys. Rev. {\bf 175},  77-98 (1968).

\bibitem{Z74} D. N. Zubarev, {\it Nonequilibrium Statistical Thermodynamics} (Consultants Bureau, New York, 1974).

\bibitem{AP81} A. I. Akhiezer and S. V. Peletminskii, {\it Methods of Statistical Physics} (Pergamon, Oxford, 1981).

\bibitem{OL79} I. Oppenheim and R. D. Levine, {\it Nonlinear transport processes: Hydrodynamics}, Physica A {\bf 99}, 383-402 (1979).

\bibitem{S14} S.-i. Sasa, {\it Derivation of Hydrodynamics from the Hamiltonian Description of Particle Systems}, Phys. Rev. Lett. {\bf 112}, 100602 (2014).

\bibitem{DLW20} J. Dufty, K. Luo, and J. Wrighton, {\it Generalized hydrodynamics revisited}, Phys. Rev. Res. {\bf 2}, 023036 (2020).

\bibitem{MG20} J. Mabillard and P. Gaspard, {\it Microscopic approach to the macrodynamics of matter with broken symmetries}, J. Stat. Mech.: Theory Exp. {\bf 2020}, 103203 (2020).

\bibitem{MG21} J. Mabillard and P. Gaspard, {\it Nonequilibrium statistical mechanics of crystals}, J. Stat. Mech.: Theory Exp. {\bf 2021}, 063207 (2021).

\bibitem{MG23} J. Mabillard and P. Gaspard, {\it Quantum local-equilibrium approach to dissipative hydrodynamics}, Phys. Rev. E {\bf 107}, 014102 (2023).

\bibitem{G22} P. Gaspard, {\it The Statistical Mechanics of Irreversible Phenomena} (Cambridge University Press, Cambridge UK, 2022).

\bibitem{F75} D. Forster, {\it Hydrodynamic Fluctuations, Broken Symmetry, and Correlation Functions} (Benjamin/Cummings, Reading MA, 1975).

\bibitem{BP76}  B. J. Berne and R. Pecora, {\it Dynamic Light Scattering} (Wiley, New York, 1976).

\bibitem{BY80}  J. P. Boon and S. Yip, {\it Molecular Hydrodynamics} (McGraw-Hill, New York, 1980)

\bibitem{G52}  M. S. Green, {\it Markoff Random Processes and the Statistical Mechanics of Time-Dependent Phenomena}, J. Chem. Phys. {\bf 20}, 1281-1295 (1952).

\bibitem{G54}  M. S. Green, {\it Markoff Random Processes and the Statistical Mechanics of Time-Dependent Phenomena. {II.}~{Irreversible} Processes in Fluids}, J. Chem. Phys. {\bf 22}, 398-413 (1954).

\bibitem{K57} R. Kubo, {\it Statistical mechanical theory of irreversible processes. {I.} {General} theory and simple applications
in magnetic and conduction problems}, J. Phys. Soc. Japan {\bf 12}, 570-586 (1957).

\bibitem{E26} A. Einstein, {\it Investigations on the Theory of the {Brownian} Movement} (E. P. Dutton \& Company, New York, 1926).

\bibitem{H60} E. Helfand, {\it Transport Coefficients from Dissipation in a Canonical Ensemble}, Phys. Rev. {\bf 119}, 1-9 (1960).

\bibitem{vH54} L. {Van Hove}, {\it Correlations in Space and Time and {Born} Approximation Scattering in Systems of Interacting Particles}, Phys. Rev. {\bf 95}, 249-262 (1954).

\bibitem{H97} J.M. Haile,  {\it Molecular Dynamics Simulation: Elementary Methods} (Wiley, New York, 1997).

\bibitem{P85} M. Pollicott, {\it On the rate of mixing of {Axiom A} flows}, Invent. Math. {\bf 81}, 413-426 (1985).

\bibitem{P86} M. Pollicott, {\it Meromorphic extensions of generalised zeta functions}, Invent. Math. {\bf 85}, 147-164 (1986).

\bibitem{R86a} D. Ruelle, {\it Resonances of chaotic dynamical systems}, Phys. Rev. Lett. {\bf 56}, 405-407 (1986).

\bibitem{R86b} D. Ruelle, {\it Locating resonances for {Axiom A} dynamical systems}, J. Stat. Phys. {\bf 44},
281-292 (1986).

\bibitem{G96} P. Gaspard, {\it Hydrodynamic modes as singular eigenstates of the {Liouvillian} dynamics: {Deterministic} diffusion}, Phys. Rev. E {\bf 53}, 4379-4401 (1996).

\bibitem{GCGD01} P. Gaspard, I. Claus, T. Gilbert, and J. R. Dorfman, {\it Fractality of the hydrodynamic
modes of diffusion}, Phys. Rev. Lett. {\bf 86}, 1506-1509 (2001).

\bibitem{KS68} L. P. Kadanoff and J. Swift, {\it Transport coefficients and the liquid-gas critical point}, Phys. Rev. {\bf 166}, 89-101 (1968).

\bibitem{G98} P. Gaspard, {\it Chaos, Scattering and Statistical Mechanics} (Cambridge
University Press, Cambridge UK, 1998).

\bibitem{LL80a} L. D. Landau and E. M. Lifshitz, {\it Statistical Physics, Part 1}, 3rd edition (Pergamon Press, Oxford, 1980).

\bibitem{B75} R. Balescu, {\it Equilibrium and Nonequilibrium Statistical Mechanics} (Wiley, New York, 1975).

\bibitem{RD77} P. R\'esibois and M. De Leener, {\it Classical Kinetic Theory of Fluids} (Wiley, New York, 1977).

\bibitem{R98} L. E. Reichl, {\it A Modern Course in Statistical Physics}, 2nd edition (Wiley, New York, 1998).

\bibitem{Z65} R. Zwanzig, {\it Time-correlation functions and transport coefficients in statistical mechanics}, Annu. Rev. Phys. Chem. {\bf 16}, 67-102 (1965).

\bibitem{BS96} L. A. Bunimovich and H. Spohn, {\it Viscosity for a periodic two disk fluid: {An} existence proof}, Commun. Math. Phys. {\bf 176}, 661-680 (1996).

\bibitem{VG03} S. Viscardy and P. Gaspard, {\it Viscosity in molecular dynamics with periodic boundary conditions}, Phys. Rev. E {\bf 68}, 041204 (2003).

\bibitem{VSG07a} S. Viscardy, J. Servantie, and P. Gaspard, {\it Transport and {Helfand} moments in the {Lennard-Jones} fluid. {I.} {Shear} viscosity}, J. Chem. Phys. {\bf 126}, 184512 (2007).

\bibitem{VSG07b} S. Viscardy, J. Servantie, and P. Gaspard, {\it Transport and {Helfand} moments in the {Lennard-Jones} fluid. {II.} {Thermal} conductivity}, J. Chem. Phys. {\bf 126}, 184513 (2007).

\bibitem{O31b} L. Onsager, {\it Reciprocal relations in irreversible processes II},  Phys. Rev. {\bf 38}, 2265-2279 (1931).

\bibitem{LL57} L. D. Landau and E. M. Lifshitz, {\it Hydrodynamic fluctuations}, JETP {\bf 5}, 512-513 (1957).

\bibitem{LL80b} L. D. Landau and E. M. Lifshitz, {\it Statistical Physics, Part 2} (Pergamon Press, Oxford, 1980).

\bibitem{OS06} J. M. {Ortiz de Z\'arate} and J. V. Sengers, {\it Hydrodynamic Fluctuations in Fluids and
Fluid Mixtures} (Elsevier, Amsterdam, 2006).

\bibitem{ED72} M. H. Ernst and J. R. Dorfman, {\it Nonanalytic dispersion relations in classical fluids. {I.} {The} hard-sphere gas}, Physica {\bf 61}, 157-181 (1972).

\bibitem{ED75} M. H. Ernst and J. R. Dorfman, {\it Nonanalytic dispersion relations for classical fluids. {II.} {The} general fluid}, J. Stat. Phys. {\bf 12}, 311-359 (1975).

\bibitem{DvBK21} J. R. Dorfman, H. {van Beijeren}, and T. R. Kirkpatrick, {\it Contemporary Kinetic Theory of Matter} (Cambridge University Press, Cambridge UK, 2021).

\bibitem{K44} N. Krylov, {\it Relaxation processes in statistical systems}, Nature {\bf 153}, 709-710 (1944).

\bibitem{S96} Ya. G. Sinai, {\it A remark concerning the thermodynamic limit of the Lyapunov spectrum}, Int. J. Bifurc. Chaos {\bf 6}, 1137-1142 (1996).

\bibitem{DP97} C. Dellago and H. A Posch, {\it {Kolmogorov-Sinai} entropy and {Lyapunov spectra} of a hard-sphere gas}, Physica A {\bf 240}, 68-83 (1997).

\bibitem{vBDPD97} H. {van Beijeren}, J. R. Dorfman, H. A. Posch, and C. Dellago, {\it Kolmogorov-Sinai entropy for dilute gases in equilibrium}, Phys. Rev. E {\bf 56}, 5272-5277 (1997).

\bibitem{SC87} Ya. G. Sinai and N. I. Chernov, {\it Ergodic properties of certains systems of 2-D discs and 3-D balls}, Russ. Math. Surveys {\bf 42}, 181-207 (1987).

\bibitem{KSS91} A. Kr\'amli, N. Sim\'anyi, and D. Sz\'asz, {\it The K-property of three billiard balls}, Ann. Math. {\bf 133}, 37-72 (1991).

\bibitem{KSS92} A. Kr\'amli, N. Sim\'anyi, and D. Sz\'asz, {\it The K-property of four billiard balls}, Commun. Math. Phys. {\bf 144}, 107-148 (1992).

\bibitem{S04} N. Sim\'anyi, {\it Proof of ergodic hypothesis for typical hard ball systems}, Ann. Henri Poincar\'e {\bf 5}, 203-233 (2004).

\bibitem{S97} R. J. Speedy, {\it Pressure of the metastable hard-sphere fluid}, J. Phys.: Condens. Matter {\bf 9}, 8591-8599 (1997).

\bibitem{R99} Y. Rosenfeld, {\it Sound velocity in liquid metals and the hard-sphere model}, J. Phys.: Condens. Matter {\bf 11}, L-71-L74 (1999).

\bibitem{AA83} W. E. Alley and B. J. Alder, {\it Generalized transport coefficients for hard spheres}, Phys. Rev. A. {\bf 27}, 3158-3173 (1983).

\bibitem{AAY83} W. E. Alley, B. J. Alder, and S. Yip, {\it The neutron scattering function for hard spheres}, Phys. Rev. A, {\bf 27}, 3174-3186 (1983).

\bibitem{W63} M. S. Wertheim, {\it Exact solution of the {Percus-Yevick} integral equation for hard spheres}, Phys. Rev. Lett. {\bf 10}, 321-323 (1963).

\bibitem{H09} D. Henderson, {\it Analytic methods for the {Percus-Yevick} hard sphere correlation functions}, Condens. Matter Phys. {\bf 12}, 127-135 (2009).

\bibitem{W62} B. P.  Welford,  {\it Note on a method for calculating corrected sums of squares and products}, Technometrics {\bf 4}, 419-420 (1962).

\bibitem{AGW70} B. J. Alder, D. M. Gass, and T. E. Wainwright, {\it Studies in Molecular Dynamics. {VIII}. {The} Transport Coefficients for a Hard-Sphere Fluid}, J. Chem. Phys. {\bf 53}, 3813-3826 (1970).

\bibitem{HCB54} J. O. Hirschfelder, C. F. Curtis, and R. B. Bird, {\it Molecular Theory of Gases and Liquids} (Wiley, New York, 1954).

\bibitem{SH03} H. Sigurgeirsson and D. M. Heyes, {\it Transport coefficients of hard sphere fluids}, Molecular Physics {\bf 101}, 469-482 (2003).

\end{thebibliography}
\end{document}